\documentclass[
 reprint,
 amsmath,amssymb,
 aps,raggedbottom
]{revtex4-2}

\usepackage{bm}
\usepackage{dcolumn}
\usepackage{graphicx}
\usepackage{mathtools}
\usepackage[T1]{fontenc}

\begin{document}

\preprint{APS/123-QED}

\title{Multifractal nonlinearity as a robust estimator of multiplicative cascade dynamics}

\author{Madhur Mangalam}
\email{Correspondence should be sent to: \href{mailto:mmangalam@unomaha.edu}{mmangalam@unomaha.edu}.}
\affiliation{Division of Biomechanics and Research Development, Department of Biomechanics, and Center for Research in Human Movement Variability, University of Nebraska at Omaha, NE 68182, USA}

\author{Aaron D. Likens}
\affiliation{Division of Biomechanics and Research Development, Department of Biomechanics, and Center for Research in Human Movement Variability, University of Nebraska at Omaha, NE 68182, USA}

\author{Damian G. Kelty-Stephen}
\email{Correspondence should be sent to: \href{mailto:keltystd@newpaltz.edu}{keltystd@newpaltz.edu}.}
\affiliation{Department of Psychology, State University of New York at New Paltz, New Paltz, New York 12561, USA}

\begin{abstract}
    Biological and psychological sciences have begun investigating whether random multiplicative cascade processes can support an organism's capacity to perceive, act, and think. Random multiplicative cascades are nonlinear processes that repeatedly divide, branch, or aggregate structures across generations, resulting in multiplicative interactions across scales that break ergodicity. Multifractal formalisms provide an apt framework to study cascades in which multifractal spectrum width $\Delta\alpha$ fluctuates depending on the number of estimable power-law relationships. Then again, multifractality without surrogate comparison can be ambiguous: the original measurement series' multifractal spectrum width $\Delta\alpha_\mathrm{Orig}$ can be sensitive to the series length, ergodicity-breaking linear temporal correlations (e.g., fractional Gaussian noise, $fGn$), or additive cascade dynamics. To test these threats, we built a suite of random cascades that differ by the length, type of noise (i.e., additive white Gaussian noise, $awGn$, or $fGn$), and mixtures of $awGn$ or $fGn$ across generations (progressively more $awGn$, progressively more $fGn$, and a random sampling by generation), and operations applying noise (i.e., addition vs. multiplication). The so-called ``multifractal nonlinearity'' $t_\mathrm{MF}$ (i.e., a $t$-statistic comparing $\Delta\alpha_\mathrm{Orig}$ and multifractal spectra width for phase-randomized linear surrogates $\Delta\alpha_\mathrm{Surr}$) is a robust indicator of random multiplicative rather than random additive cascade processes irrespective of the series length or type of noise. $t_\mathrm{MF}$ is more sensitive to the number of generations than the series length. Furthermore, the random additive cascades exhibited much stronger ergodicity breaking than all multiplicative analogs. Instead, ergodicity breaking in random multiplicative cascades more closely followed the ergodicity-breaking of the constituent noise types---breaking ergodicity much less when arising from ergodic $awGn$ and more so for noise incorporating relatively more correlated $fGn$. Hence, $t_\mathrm{MF}$ is a robust multifractal indicator of multiplicative cascade processes and not spuriously sensitive to ergodicity breaking, suggesting continued utility in studying the emergence and creativity of biological and psychological behavior.
\end{abstract}


\maketitle

\section{Introduction}

\subsection{Cascade-like interactivity in biological and psychological behavior}

Biological and psychological sciences over the past century have witnessed an ongoing and creative tension between two classes of metaphors---both, curiously, promoted by the same creative scientist, namely, Alan Turing \cite{kelso1995dynamic}. Turing first provided the biological and psychological sciences with the useful ``computer'' metaphor of the mind and brain, fostering decades of empirical research. This approach has been invaluable for attempts to decompose biological and psychological functions into independent constituent processes for a linear or linearized model, usually endowed with hierarchical structure and Bayesian support \cite{de2018expectations,simon1970sciences}. The vast popularity of the computer metaphor has long overshadowed the other metaphor that he promoted---that of the ``cascade.'' Although Turing was not the first to compare the mind to the turbulent nonlinearity of fluid flow \cite{james1890stream,richardson1930analogy}, his 1952 work on morphogenesis pioneered the attempt to make cascading dynamics a computationally tractable question for biological sciences \cite{turing1952chemical} with potential applications to how biological systems organized themselves to support thought \cite{dawson2007essential}. \textit{Cascades} are nonlinear processes that repeatedly divide, branch, or aggregate structures across generations, enacting or reflecting nonlinear interactions across scales \cite{mandelbrot1982fractal,turcotte2002self,turing1990chemical}. A rapidly expanding corpus of research demonstrates that cascade instability supports perceiving, acting, and thinking in a wide variety of organisms (\cite{dixon2012multifractal,kelty2013gaze,koorehdavoudi2017multi,rezania2021multifractality,roeske2018multifractal,schmitt2001multifractal,seuront2004random,seuront2014anomalous,stephen2011strong,wawrzkiewicz2020multifractal}; see \cite{kelty2022turing,ihlen2013multifractal} for reviews). Turing's cascade instability suggests a geometrical framework governed by power laws that can be studied using multifractal formalisms \cite{ihlen2010interaction,kantelhardt2002multifractal,lovejoy2018weather,mandelbrot1974intermittent}. Cascades have facilitated an immensely useful discourse on the creative tension between the computer-like rule following amongst constituent biological and psychological phenomena and their more rate-dependent fluid-like dynamics of spanning multiple space and time scales \cite{pattee2013epistemic}.

The tension has empirically run aground on the theoretical issue of ergodicity. In short, the linear or linearized modeling that can effectively reduce biological and psychological functions to independent constituent processes depends on an assumption of ergodicity \cite{molenaar2008implications,molenaar2009analyzing,molenaar2009new}. Raising novel challenges to the linear model so effective at giving voice to the computer metaphor is the fairly widespread empirical fact that short of a limited set of well-behaved model systems with discernibly quiescent scales or modes (e.g., \cite{mcleish2015there}), much of the empirical biological and psychological records are ergodicity breaking \cite{colombo2021non,kelty2022fractal,kelty2023multifractaldescriptors,mangalam2021point,mangalam2022ergodic,mangalam2023ergodic}. Cascades are widely expected to generate ergodicity-breaking dynamics \cite{hu2016dynamics,li2022non,weigel2011ergodic}, and to the extent that cascades generate multifractal statistics \cite{mandelbrot1974intermittent}, we can often find multifractal systems embodying ergodicity-breaking \cite{de2020multifractality,mace2019multifractal,mahmoodi2020dynamical}. Specifically, in contrast to literature that has long waxed poetically about stream-like context sensitivities of the body or mind, multifractal modeling allows us quite soberly to estimate a quantitative estimate of how strongly events may interact across scales when observations fail to fit the structures of the classic linear model. Multifractal formalisms estimate a spectrum $f(\alpha)$ of singularity strengths $\alpha$ controlling power-law growth in relation to the measurement scale. The most commonly used feature---the multifractal spectrum width, $\Delta\alpha$, fluctuates depending on the number of estimable power-law relationships and characterizes the heterogeneity attributable to nonlinear interactions across scales due to multiplicative cascade dynamics \cite{ihlen2012introduction}. Not only is multiplicative cascade dynamics likely to break ergodicity, but estimates of multifractal features in ergodicity-breaking series may provide the most reliable descriptors least likely to break ergodicity themselves \cite{kelty2022fractal,kelty2023multifractaldescriptors,mangalam2021point,mangalam2022ergodic,mangalam2023ergodic}. Hence after a sluggish start in keeping pace with the computer metaphor of the brain and mind, the cascade metaphor is gradually catching its explanatory stride. It may explain both the ergodicity breaking in plain, widespread evidence and entails a central role for multifractal formalism in arriving at reliable, representative descriptors that can support stable modeling of cause and effect relationships. The present work focuses on effectively constructing a continuous measure of nonlinear correlations for empirically modeling cascade-like behaviors.

\subsection{Estimating nonlinear interactions across scales due to multiplicative cascade dynamics in terms of multifractal nonlinearity}

\subsubsection{Theoretical background of ``multifractal nonlinearity''}

Ultimately, interpreting multifractal evidence as indicating nonlinearity requires surrogate testing---a process that compares the multifractal spectrum of empirical series to that of synthetic data that (i) takes the measured values, (ii) randomizes their phase, and (iii) preserves their linear correlations. \cite{lancaster2018surrogate,mandic2008characterization,schreiber1997discrimination,schreiber2000surrogate,theiler1992testing}. Linear correlations in empirical series can artificially raise $\Delta\alpha$, independently or in addition to nonlinear correlations \cite{chen2016finite,zhou2012finite}. Therefore, it is customary to compare the multifractal spectrum width of the original series, $\Delta\alpha_\mathrm{Orig}$, to that of a set of surrogates with matching mean, variance, and autocorrelation function, $\Delta\alpha_\mathrm{Surr}$. The preferred surrogate is obtained using the Iterative Amplitude Adjusted Fourier Transform (IAAFT) algorithm, which closely matches the distribution and the power spectrum in the original data and the surrogates by iteratively replacing Fourier amplitudes with the correct values and scaling the distribution \cite{schreiber1996improved}. Nonzero $\Delta\alpha_\mathrm{Surr}$ indicates that $\Delta\alpha_\mathrm{Orig}$ is inflated due to skewed histograms of fluctuations \cite{struzik2004econophysics}.

There are, to date, at least two ways to establish nonlinearity in multifractal tests. The choice between these two approaches exemplifies more general choices that empirical science has long entertained between dichotomous null-hypothesis testing and more continuous ways of expressing effect size \cite{cohen2016earth}. In the former option, we can establish a dichotomous notion of nonlinearity by testing the null hypothesis of linearity \cite{theiler1992testing}. This strategy involves selecting a preferred Type I error rate and building as many linear surrogates as required to evaluate the original series's relative extremity of the multifractal spectrum width. For instance, when assessing the nonlinearity of a series at $p<0.05$, we generate $20$ surrogates and reject the null hypothesis of the original series’ linearity if it has a wider multifractal spectrum than $19$ of these surrogates---choosing a lower Type I error rate (e.g., $p<0.01,0.005,0.001$) entails generating a larger sample of surrogates (e.g., $100,200,1000$, respectively). We will reject the null hypothesis if the original series shows a more extreme multifractal spectrum than $99,199,999$ surrogates, respectively. This dichotomous approach of rejecting the null hypothesis of linearity faces the long-appreciated challenges of interpreting $p$-values with little clear guidance as to what effect size would be informative. A likely candidate is the indicator of standardized difference, Cohen's $d$: $\frac{\Delta\alpha_\mathrm{Orig}-\overline{\Delta\alpha_\mathrm{Surr}}}{\sigma_{\Delta\alpha_\mathrm{Surr}}}$. However, there is no reason to expect that surrogate spectrum widths meet Cohen's $d$ assumption of Normality of empirical series (e.g., \cite{wooff2013robust}).

We recommend the $t$-statistic comparing the original multifractal spectrum width to surrogate spectra widths for a continuous measure of multifractal nonlinearity. Here, we expressly intend this term ``multifractal nonlinearity'' to mean the $t$-statistic realizing this comparison, which we abbreviate elsewhere as $t_\mathrm{MF}$. Despite dependence on standard error rather than standard deviation, using $t$ as an effect size is not without precedence (e.g., \cite{kendler2018prediction,yeung2016endogenous}). Despite the disadvantage of being dependent on (the square root of) the sample size of surrogates \cite{fan2020fdi,tench2017coordinate}, the $t$-statistic helps us pursue a continuous complement to the dichotomy of a null-hypothesis rejection \cite{kelty2023multifractaltest}. This $t$-statistic increases when the original series’ multifractality diverges more and more from the multifractality attributed to the linear structure of IAAFT surrogates. Indeed, modelers wishing to use the $t$-statistic to estimate a $p$-value can do so. Although no fixed minimum sample size exists for a $t$-test, $32$ surrogates yield a suitable $t$-statistic given the conditions of Type I error of $p<0.05$, ``medium'' effect size of $0.5$, and power of $0.5$. Notably, though, we are advocating using the $t$-statistic in terms of its continuous value---not as a means to get a $p$-value. The continuous $t$-statistic values serve as effective predictors of psychological results for outcomes classed as perception-action \cite{bell2019non,carver2017multifractal,doyon2019multifractality,harrison2014multiplicative,jacobson2021multifractality,kelty2018multifractal,kelty2021multifractal,kelty2023multifractalnonlinearity,mangalam2020multifractal,mangalam2020multiplicative,palatinus2014haptic,stephen2012multifractal} and cognitive \cite{bloomfield2021perceiving,booth2018expectations,dixon2012multifractal,ward2018bringing}.

\subsubsection{Critiques and questions about clarifying the empirical relationships between ``multifractal nonlinearity'' $t_\mathrm{MF}$}

The present work aims to clarify the empirical connection between multiplicative interactions across scales and multifractal nonlinearity. In particular, by ``multifractal nonlinearity'' we now mean a $t$-statistic $t_\mathrm{MF}$ comparing $\Delta\alpha_\mathrm{Orig}$ and multifractal spectra width for phase-randomized linear surrogates $\Delta\alpha_\mathrm{Surr}$, this estimate of nonlinearity could be a robust indicator of random multiplicative rather than random additive cascade dynamics, irrespective of the series length or type of noise. The challenge with multifractal investigations is that a nonzero $\Delta\alpha$---more than one estimable power-law relationship---can be opaque, with $t_\mathrm{MF}$ potentially at risk for inheriting and algorithmically adding more opacity. The clarity with which we may assert that random multiplicative cascades entail multifractal statistics does not amount to the clarity of a single multifractal spectrum estimated from empirical data; that is, a nonzero $\Delta\alpha$ does not necessarily imply nonlinear interactions across scales due to multiplicative cascade dynamics \cite{veneziano1995multifractal}, as some linear models can also generate a variety of power-law scaling in behavior and cognition \cite{shlesinger1987levy,wagenmakers2004estimation}. For present purposes, we wish to test whether successive generations of random multiplicative interactions across scales increase $t_\mathrm{MF}$ and whether this relationship was strong compared to linear effects on $t_\mathrm{MF}$.

Indeed, a class of critiques raises concerns about the usefulness of multifractal formalisms instead of algorithmically ``simpler'' methods for modeling the interactivity and ergodicity breaking of biological and psychological variability. Each would make interpreting $t_\mathrm{MF}$ simply as evidence of repeated generations of multiplicative interactions across scales increasingly difficult. First, the capacity of monofractality to vary unsystematically with finite sample sizes could entail length spuriously increasing a multifractal estimate of what might sooner be a monofractal process (e.g., \cite{ihlen2010interaction}). So, the risk of false-positive differences from surrogates makes $t_\mathrm{MF}$ less informative about interactions across scales. Second, fractional Gaussian noise ($fGn$) can break ergodicity \cite{deng2009ergodic,mangalam2022ergodic}, so perhaps the appearance of ``interactivity'' explicitly as multiplicative noise across scales was an unnecessary gloss borrowed from poetic rhetoric, and perhaps the best explanation for ergodicity breaking is the $fGn$ suffusing mind and body. Hence, patterns in $t_\mathrm{MF}$ may sooner reflect ergodicity breaking than nonlinear interactions across scales due to multiplicative cascade dynamics. Third, $fGn$ can arise from cascade-like fractionation using only additive rather than multiplicative noise added at each generation, that is, from random additive cascades sooner than random multiplicative ones \cite{taylor2021prime}. So, the multifractality estimated from unsystematically varying and poorly sampled monofractality could reflect a random additive cascade sooner and not reflect any of the nonlinearity of multiplicative noise across scales---many of which factors could spuriously increase $t_\mathrm{MF}$ without building in any nonlinear interactions across scales. These foregoing points spell out a case in which multifractal formalisms are excessive. On the other hand, it is possible that all of these contingencies are not threats to the interpretation of multifractal spectra for inferring random multiplicative cascade processes. That is, above and beyond such additive and monofractal sources of multifractal variance, increases in multifractal spectrum width might reflect successive generations of multiplicative cascade dynamics, and random multiplicative cascades may be the stronger source of ergodicity breaking.

\subsection{The missing link between multifractality and multifractal nonlinearity}

Multifractal nonlinearity $t_\mathrm{MF}$ may provide a more comprehensive, continuous way to assess nonlinear interactions across scales due to multiplicative cascade dynamics and confers several advantages over the dichotomous approach of rejecting the null hypothesis of linearity. However, the relationship of $t_\mathrm{MF}$ with the strictly linear and strictly nonlinear sources of multifractality has remained elusive. Except for a few theoretical simulations \cite{lee2017cascade,mangalam2022ergodic,kelty2023multifractaldescriptors,kelty2022fractal}, the use of $t_\mathrm{MF}$ has been demonstrated more often in empirical works. Given the repeated efficacy of $t_\mathrm{MF}$ as a predictor of empirical data, it is now important to ensure the validity of $t_\mathrm{MF}$ for representing multiplicative interactions across scales due to multiplicative cascade dynamics and not spuriously sensitive to other factors like length or ergodicity breaking. How might $t_\mathrm{MF}$ depend on the data collection process, such as the length of measurement series, noise from various exogenous sources, and the data processing steps, such as time-series segmentation? Do different data collection and processing practices influence multifractality and multifractality differently? According to the foregoing reasoning about random multiplicative cascades and the known sensitivity of nonzero multifractality to the presence of multiplicative cascading processes, we might begin this investigation of the test comparing multifractal spectra to surrogates' multifractal-spectral widths with two substantive predictions: First, $t_\mathrm{MF}$ might well be greater for random multiplicative cascades than for random additive cascades that do not enact the nonlinear interactions across scales. Second, progressively more interactions across scales (i.e., across ever more generations of a cascade) might accentuate multifractal nonlinearity further, yielding greater $t_\mathrm{MF}$ with greater numbers of generations. A necessary caution for the second prediction is that the number of generations will often covary with length. So, we must test this idea with the traditional binomially-fractured cascade models and cascades designed to maintain the same length but reflect progressively more interactions across timescale. For instance, does a wider multifractal spectrum always imply stronger nonlinear interactions across scales due to multiplicative cascade dynamics? Depending on the noise surrounding the measured process, some noise will induce sequences that break ergodicity (e.g., as we find in long-range correlated $fGn$). Prior work has shown that cascades can break ergodicity, so $t_\mathrm{MF}$ may be spuriously sensitive to ergodicity breaking. These missing links between multifractality and multifractal nonlinearity make it challenging to assess how much of the observed multifractality can be attributed to nonlinear interactions across scales due to multiplicative cascade dynamics. The present study investigates these issues with surrogate testing: the relationship between multifractality, $\Delta\alpha_\mathrm{Orig}$, and multifractal nonlinearity, $t_\mathrm{MF}$, as it pertains to strictly linear and strictly nonlinear sources of multifractality.

We tested the following four hypotheses, considering various data collection and processing choices the behavioral scientist might face while performing multifractal analysis. Hypotheses 1 and 2 both encode the predictions about how $t_\mathrm{MF}$ might serve to infer multiplicative cascading processes even above and beyond the earlier hypothesized effects. Hypothesis 3 encodes our predictions of possible spurious inflation of $t_\mathrm{MF}$ due to length or linear factors known to break ergodicity. 

\textbf{Hypothesis 1:} We hypothesized that random multiplicative cascades would show greater multifractal-spectrum width $\Delta\alpha(q)_\mathrm{Orig}$.

\textbf{Hypothesis 2A:} We hypothesized that the effect of multiplicativity would lead to greater differences of $t_\mathrm{MF}$ from zero as the random multiplicative cascade enact progressively more multiplicative interactions generations, above and beyond the influence of the increasing series length across successive generations inherent to the cascade model. We expected progressively more generations of random multiplicative interactions across scales to produce cascades with both greater multifractal nonlinearity (i.e., greater $t_\mathrm{MF}$) and more frequent significant differences from zero (i.e., $t_\mathrm{MF}>2.04$). However, the difficulty here is the confounding of length with the number of generations in a cascade model that hinges on binomially fracturing fewer parent cells in one generation into more descendant cells in later generations. Therefore, this hypothesis warranted subsequent corollary hypotheses. 

\textbf{Hypothesis 2B:} We hypothesized that a longer cascading series would elicit greater differences of $t_\mathrm{MF}$ from zero. Short measurement series reflecting cursory observation of ongoing multiplicative cascade processes might exhibit nonlinear correlations reflecting multiplicative interactions across scales. However, the shorter length admits fewer possible phase-randomized sequences, making any given phase randomization more likely to resemble the original phase relations. In contrast, the greater series length naturally increases the number of possible phase-randomized surrogates that could satisfy the same linear correlations. Hence, greater length allows for more variety in phase randomization and a greater likelihood for genuine nonlinearity to appear. However, the evidence from $t_\mathrm{MF}$ might be less valuable indeed if the greater length of any process were enough to inflate the multifractal-spectrum width $\Delta\alpha_\mathrm{Orig}$, for example, as though multifractality were no more than unsystematic error collected through measuring an $fGn$ series. Length dependence of $t_\mathrm{MF}$ might be more informative if it were not $\Delta\alpha_\mathrm{Orig}$ that were changing but if, instead, greater length leading to narrower multifractal spectra for surrogates. This latter case of length-dependent $t_\mathrm{MF}$ would instead entail that the original series was not reducible to $fGn$. Instead, the longer observation of the linear process reduces to more clearly a random process like $fGn$ with a more monosingular structure.

\textbf{Hypothesis 2C:} Our hypothesis centered on the notion that the impact of multiplicativity would result in more pronounced deviations of $t_\mathrm{MF}$ from zero, especially as the multiplicative cascade advanced through successive generations. We aimed to isolate this effect from the natural progression of the cascade model, which inherently extended the series length across generations. To empirically assess this hypothesis, our plan involved constructing a cascade model with a series of fixed lengths, wherein the values could be repeated over extended scales. This repetition would enable subsequent generations to operate exclusively on progressively smaller scales. Incorporating repeated values within the series would be extensive enough to facilitate similar interactions across scales, effectively simulating the advancement of generations without necessitating an increase in the number of values in each subsequent generation.

\textbf{Hypothesis 3:} We hypothesized that (i) cascades arising through the application of $fGn$---in pure form and also various mixtures of $fGn$ with $awGn$---at each generation would yield original sequences more likely to break ergodicity than cascades using pure $awGn$ and so (ii) a greater tendency to break ergodicity in $fGn$-driven cascades than $awGn$-driven might explain differences of $t_\mathrm{MF}$ from zero across the different cascades and noise types.

\section{Methods}

\subsection{Generating random cascade series}

\subsubsection{Binomial fracturing and binomial noise terms in prior cascade simulation}

\begin{figure*}
    \includegraphics[width=6.125in]{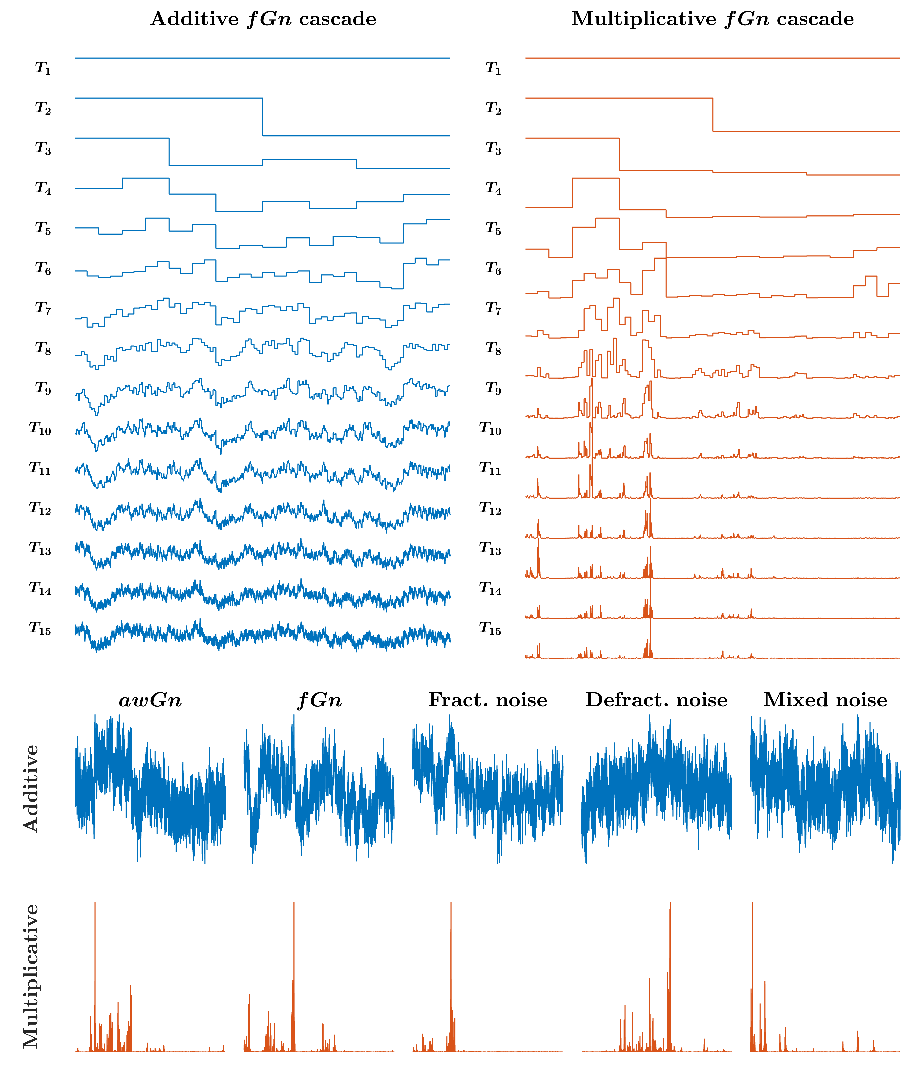}
    \caption{A random additive or multiplicative cascade serves as a mathematical framework for elucidating how the distribution of quantities or events can evolve across increasingly smaller sample sizes and shorter time intervals. In the \textit{upper-left} quadrant, we observe a random additive cascade characterized by $fGn$ noise terms manifesting across $14$ generations of iterative splitting and additive interactions. Meanwhile, in the \textit{upper-right} quadrant, a random multiplicative cascade with $fGn$ noise terms unfolds over $14$ generations, showcasing multiplicative interactions. Each curve has been normalized by its maximum value and vertically shifted by $1$ unit to enhance clarity. The \textit{lower panel} presents a compelling display of both cascade types, exemplifying variations in interactivity that span from additive to multiplicative. Both cascade types encompass noise types, including $awGn$, $fGn$, \textit{fractalizing}, \textit{defractalizing}, and \textit{mixed}. Observe the pronounced prevalence of exceedingly large events in random multiplicative cascades, not additive cascades.}
    \label{fig: f1}
\end{figure*}

Cascades are iterative manipulations of $n_{g}$ cells over $g$ generations spreading the proportion $p_{i,j}$ contained in each $i$th parent cell in generation $j$ (for $j<g,i\leq n_{j}$) across $n_{c}$ children cells in generation $j+1$ each containing $p_{k,j+1}$, for $i\leq k \leq n_{j+1}$. We produced $10$ types of cascade series noted below in more detail, all of which were binomial cascades. Cascades are ``binomial'' when each parent cell bequeaths proportions to two children cells in the next generation (i.e., $n_{c}=2$). Half of the cascades were multiplicative, and half were additive. Cascades are multiplicative or additive when the $n_{c}$ children-cell proportions reflect $n_{c}$ different multiplications or additions, respectively, governing the distribution of the parent cell’s proportion. Our cascades were binomially fractured at each generation. In the binomially-fractured cascade process, it is customary to use binomial noise terms to match the pairing of children cells (Fig.~\ref{fig: f1}). For every $i$th parent cell in the $j$th generation, applying these binomial noise terms could follow a deterministic method, for example, applying the same noise terms $W_{1}=0.25$ and $W_{2}=0.75$ to calculate the proportions in the $2i-1$th and the $2i$th children cells in the $j+1$th generation as $p_{2i-1,j+1}=p_{i,j}\cdot W_{1}$ and $p_{2i,j+1}=p_{i,j} \cdot W_{2}$, respectively.

\subsubsection{Beyond binomial noise terms: Binomial-fracturing cascades with noise terms defined across entire generations to test for effects of length, $fGn$, and multiplicativity}

To understand the multifractal descriptors of empirical data, we use numerical simulations of random cascades resembling the heterogeneity of random component processes and measured outcomes from biological and psychological sciences. The present work aims to maintain the binomial fracturing from parent to children cells but to define the noise terms randomly beyond the binomial ($W(1)$,$W(2)$) structure. It is certainly possible to randomize the binomial multiplicative cascade, for instance, by drawing $W_{1}$ randomly from a uniform distribution from $(0.25,0.75)$ and calculating $W_{2}=1-W_{1}$ \cite{kelty2023multifractaldescriptors}. However, given our concerns about $fGn$ and its potential capacity to vary unsystematically with greater length and break ergodicity, we wanted to define the noise terms beyond strictly binomial specification. Specifically, we maintained a two-children fracturing of parent cells in subsequent generations to maintain the same rate of lengthening the cascading series. The major difference in what follows is that we generated the cascade series by generating noise terms defined across the entire generation, either as $awGn$, $fGn$, or gradual accentuation or diminution of temporal correlations in $awGn$ and $fGn$, respectively. That is, rather than defining $W_{1}$ and $W_{2}$ separately for each $i$th parent as in previous simulations \cite{kelty2023multifractaldescriptors}, we defined the noise term $W_{t,j+1}, t=1,2,\dots,n_{j+1}$-length series drawn from a Gaussian distribution with $\mu=1$ and $\sigma=1$ with dependence of noise terms in $fGn$ from $i=1$ to $i=n_{j+1}$ determined by $H_\mathrm{fGn}$ which was $0.5$ or $1$ for ideal $awGn$ or ideal $fGn$, respectively.

In these terms, we can now define the $10$ types of cascades we simulated. All cascades are binomially fracturing (i.e., with two children in $j+1$th generation for each $i$th parent in the $j$th generation), and they all specify noise terms across entire generations of children. They differ in three dimensions: additivity or multiplicativity, noise type ($awGn$ or $fGn$), and transitions (or not) between noise types across generations. Whereas the first two dimensions are straightforward dichotomies, we have defined the transition between noise types across generations in three ways. First, in what we have called the ``fractalizing'' cascade, we begin with anticorrelated Gaussian noise defined as $fGn, H_\mathrm{fGn}=0$ and gradually increase $H_\mathrm{fGn}$ to $1$ across $15$ generations. Second, in what we have called the ``defractalizing'' cascade, we begin with ideal $fGn$ defined as $H_\mathrm{fGn}=1$ and gradually decrease $H_\mathrm{fGn}$ to $0$ across $15$ generations. Third, in what we have called the ``mixed'' cascade, we randomly select either ideal $fGn$ and ideal $awGn$ at each generation.

\begin{enumerate}
    \item \textbf{\textit{Additive}} \boldmath$awGn\;$\unboldmath \textbf{cascades.} Noise terms for generation $j+1$ follow $W_{t,j+1}=awGn,\mu =1,\sigma =1$ (where $\mu$ is mean and $\sigma$ is standard deviation), and the $2i-1$th and $2i$th children cells in the $j+1$th generation holds proportion $p_{i}+W_{2i-1,j+1}$ and $p_{i}+W_{2i,j+1}$, respectively.
    \item \textbf{\textit{Additive}} \boldmath$fGn\;$\unboldmath \textbf{cascades.} Noise terms for generation $j+1$ follow $W_{t,j+1}=fGn,\mu =1,\sigma =1,H_\mathrm{fGn}=1$, and the $2i-1$th and $2i$th children cells in the $j+1$th generation holds proportion $p_{i}+W_{2i-1,j+1}$ and $p_{i}+W_{2i,j+1}$, respectively.
    \item \textbf{\textit{Additive fractalizing noise cascades.}} Noise terms for generation $j+1$ follow $W_{t,j+1}=fGn,\mu =1,\sigma = 1,H_\mathrm{fGn}=\frac{1}{14}\cdot (j+1)$, and the $2i-1$th and $2i$th children cells in the $j+1$th generation holds proportion $p_{i}+W_{2i-1,j+1}$ and $p_{i}+W_{2i,j+1}$, respectively.
    \item \textbf{\textit{Additive defractalizing noise cascades.}} Noise terms for generation $j+1$ follow $W_{t,j+1}=fGn,\mu =1,\sigma =1,H_\mathrm{fGn}=1-\frac{1}{14}\cdot (j+1)$, and the $2i-1$th and $2i$th children cells in the $j+1$th generation holds proportion $p_{i}+W_{2i-1,j+1}$ and $p_{i}+W_{2i,j+1}$, respectively.
    \item \textbf{\textit{Additive mixed noise cascades.}} Noise terms for $j+1$ follow $W_{t,j+1}=fGn,\mu =1,\sigma =1,H_\mathrm{fGn}=\{\frac{0.5}{1}$, and the $2i-1$th and $2i$th children cells in the $j+1$th generation holds proportion $p_{i}+W_{2i-1,j+1}$ and $p_{i}+W_{2i,j+1}$, respectively.
    \item \textbf{\textit{Multiplicative}} \boldmath$awGn\;$\unboldmath \textbf{cascades.} Noise terms for generation $j+1$ follow $W_{t,j+1}=awGn,\mu =1,\sigma =1$, and the $2i-1$th and $2i$th children cells in the $j+1$th generation holds proportion $p_{i}\cdot W_{2i-1,j+1}$ and $p_{i}\cdot W_{2i,j+1}$, respectively.
    \item \textbf{\textit{Multiplicative}} \boldmath$fGn\;$\unboldmath \textbf{cascades.} Noise terms for generation $j+1$ follow $W_{t,j+1}=fGn,\mu =1,\sigma =1,H_\mathrm{fGn}=1$, and the $2i-1$th and $2i$th children cells in the $j+1$th generation holds proportion $p_{i}\cdot W_{2i-1,j+1}$ and $p_{i}\cdot W_{2i,j+1}$, respectively.
    \item \textbf{\textit{Multiplicative fractalizing noise cascades.}} Noise terms for generation $j+1$ follow $W_{t,j+1}=fGn,\mu =1,\sigma = 1,H_\mathrm{fGn}=\frac{1}{14}\cdot (j+1)$, and the $2i-1$th and $2i$th children cells in the $j+1$th generation holds proportion $p_{i}\cdot W_{2i-1,j+1}$ and $p_{i}\cdot W_{2i,j+1}$, respectively.
    \item \textbf{\textit{Multiplicative defractalizing noise cascades.}} Noise terms for generation $j+1$ follow $W_{t,j+1}=fGn,\mu =1,\sigma =1,H_\mathrm{fGn}=1-\frac{1}{14}\cdot (j+1)$, and the $2i-1$th and $2i$th children cells in the $j+1$th generation holds proportion $p_{i}\cdot W_{2i-1,j+1}$ and $p_{i}\cdot W_{2i,j+1}$, respectively.
    \item \textbf{\textit{Multiplicative mixed noise cascades.}} Noise terms for $j+1$ follow $W_{t,j+1}=fGn,\mu =1,\sigma =1,H_\mathrm{fGn}=\{\frac{0.5}{1}$, and the $2i-1$th and $2i$th children cells in the $j+1$th generation holds proportion $p_{i}\cdot W_{2i-1,j+1}$ and $p_{i}\cdot W_{2i,j+1}$, respectively.
\end{enumerate}

These $10$ binomial cascades covered Hypotheses 1 through 4. We generated cascades of $2^{14}$ samples in the $15$th and final generation with the explicit intention to explain what might drive multifractal patterns in biology and psychology. We simulated a total of $100$ cascading series for each type. 

\subsubsection{Padding cascades with consecutive repetitions of cell values to disentangle generation number from length}

Hypothesis 2C motivated a slightly different set of cascades than Hypothesis 2A and 2B. Specifically, Hypothesis 2A and 2B predict that more generations and longer series will increase the likelihood that random multiplicative cascades will show greater multifractal nonlinearity $t_\mathrm{MF}$. However, Hypothesis 2C predicts that the number of successive generations is an important source of multifractal nonlinearity irrespective of changes in length. That is to say, Hypothesis 2B only respects the fact that greater length leaves more room for the possibility of unsystematic variation in fractal scaling, as well as more risk of spurious nonzero values of $t_\mathrm{MF}$. Hypothesis 2C addresses the possibility that $t_\mathrm{MF}$ may be sensitive to the number of parent-to-child-cell generations supporting the cascade process. The challenge is that the standard binomial cascades conflate generations with length because the fracturing of parent cells necessarily doubles the length. Hence, we generated a second set of cascades to disentangle length from several generations. Specifically, using the $9$th through $15$th generation of the original cascades, we padded each generation of each cascade series with repeated values to have the same length, that is, $2^{14}$. We can illustrate what we mean in short form as follows. The traditional deterministic cascade process with a $25\%$--$75\%$ bias across the child cells might yield the following three generations:
\begin{itemize}
    \item Generation $1$: $1$
    \item Generation $2$: $0.25$, $0.75$
    \item Generation $3$: ($0.25\times 0.25=$) 0.0625, ($0.25\times 0.75=$) 0.1875, ($0.75\times 0.25=$) 0.1875, ($0.75\times 0.75=$) 0.5625.
\end{itemize}
Instead, this strategy of padding the cascade series to have the same length throughout generations while also reflecting successive interactions of generations across scales would yield the following:
\begin{itemize}
    \item Generation $1:\;1,1,1,1$
    \item Generation $2:\;0.25,0.25,0.75,0.75$
    \item Generation $3:\;0.0625,0.1875,0.1875,0.5625$
\end{itemize}
In the latter cases, the important difference from the traditional cascade process is that our padded cascades have the same length in each generation. However, in both cases, the generations reflect the applications of noise to progressively shorter scales. That is, the profile of disparity across the cascade series remains comparable between each $j$th generation of the traditional cascade and the corresponding $j$th generation of the padded cascade. For instance, in the second generation, the first half of the cascade series in both cases has a value of $0.25$. The second half has a value of $0.75$, reflecting a prior application of noise terms to a span of $l/2$, and the third generation reflects the exact application of noise terms to a span of $l/4$. However, the length remains the same in the padded cascades at each generation.

More formally, we describe what this padding process would look like for the first three generations of a cascade after the $8$th generation: the first generation would have $2^{14}$ instances of the same value; the second generation would have the first child cell's value repeated for the $1$st through $2^{13}$th cells and the second child cell's value repeated for the $2^{13}+1$st cell through $2^{14}$th cell; and the third generation would consist of the first $2^{12}$ consecutive cells all containing the first third-generation child cell's value, the second $2^{12}$ consecutive cells all containing the second third-generation child cell's value, the third $2^{12}$ consecutive cells all containing the third third-generation child cell's value, and the last $2^{12}$ consecutive cells all containing the fourth third-generation child cell's values. We similarly padded the values of the surrogate series to produce the same $2^{14}$-sample length. For what follows, we only designed these padded sequences to enforce $l= 2^{14}$ for the $9$th through the $15$th generations because we wanted to instill the present work the utmost confidence in the original series having reliable IAAFT surrogates using $l\leq 2^{9}$.

For each numerically simulated series, we calculated the multifractal spectrum width $\Delta\alpha$ according to the Chhabra and Jensen's \cite{chhabra1989direct} direct method, the $t$-statistic $t_\mathrm{MF}$ comparing multifractal-spectrum width $\Delta\alpha_\mathrm{Orig}$ of the original series to a sample of surrogates' multifractal spectrum widths $\Delta\alpha_\mathrm{Surr}$, and Thirumalai-Mountain metric of ergodicity breaking $EB$ for original series as well as for randomized shufflings of the original series. Whereas the calculation of $t_\mathrm{MF}$ draws on phase-randomized surrogates to compare original series, evaluating $EB$ involves comparison to randomized shufflings because shuffling destroys both linear and nonlinear correlational sources of ergodicity breaking.

\subsection{Multifractal analysis}

\subsubsection{Assessing multifractal nonlinearity using the direct-estimation of singularity spectrum}

We used Chhabra and Jensen’s \cite{chhabra1989direct} direct method for all analyses. This method estimates multifractal spectrum width $\Delta\alpha$ by sampling a series $x(t)$ at progressively larger scales using the proportion of signal $P_{i}(n)$ falling within the $v$th bin of scale $n$ as
\begin{equation*}
    P_{v}(n)=\frac{\sum \limits_{k=(v-1)\,n+1}^{v\cdot N_{n}}x(k)}{\sum{x(t)}}, \quad n=\{4,8,16,\dots \}<T/8. \tag{1}\label{eq: 1}
\end{equation*}
As $n$ increases, $P_{v}(n)$ represents a progressively larger proportion of $x(t)$,
\begin{equation*}
    P(n)\propto n^{\alpha}, \tag{2}\label{eq:2}
\end{equation*}
suggesting a growth of the proportion according to one ``singularity'' strength $\alpha$ \cite{mandelbrot1982fractal}. $P(n)$ exhibits multifractal dynamics when it grows heterogeneously across time scales $n$ according to multiple singularity strengths, such that
\begin{equation*}
    P(n_{v})\propto n^{\alpha_{v}}, \tag{3}\label{eq: 3}
\end{equation*}
whereby each $v$th bin may show a distinct relationship of $P(n)$ with $n$. The width of this singularity spectrum, $\Delta\alpha=(\alpha_\mathrm{max}-\alpha_\mathrm{min})$, indicates the heterogeneity of these relationships \cite{halsey1986fractal,mandelbrot2013fractals}.

Chhabra and Jensen's \cite{chhabra1989direct} method estimates $P(n)$ for $N_{n}$ nonoverlapping bins of $n$-sizes and transforms them into a ``mass'' $\mu(q)$ using a $q$ parameter emphasizing higher or lower $P(n)$ for $q>1$ and $q<1$, respectively, in the form
\begin{equation*}
    \mu_{v}(q,n)=\frac{\bigl[P_{v}(n)\bigl]^{q}}{\sum\limits_{j=1}^{N_{n}}\bigl[ P_{j}(n)\bigl]^{q}}. \tag{4}\label{eq: 4}
\end{equation*}
Then, $\alpha(q)$ is the singularity for mass $\mu$-weighted $P(n)$ estimated as
\begin{equation*}
    \alpha(q)=-\lim_{N_{n}\to \infty}\frac{1}{\ln N_{n}}\sum_{v=1}^{N_{n}}\mu_{v} (q,n)\ln P_{v}(n)
\end{equation*}
\begin{equation*}
    = \lim_{n\to 0}\frac{1}{\ln n}\sum_{v=1}^{N_{n}}\mu_{v}(q,n)\ln P_{v}(n). \tag{5}\label{eq: 5}
\end{equation*}
Each estimated value of $\alpha(q)$ belongs to the multifractal spectrum only when the Shannon entropy of $\mu(q,n)$ scales with $n$ according to the Hausdorff dimension $f(q)$ \cite{chhabra1989direct}, where
\begin{equation*}
    f(q)=-\lim_{N_{n}\to \infty}\frac{1}{\ln N_{n}} \sum_{v=1}^{N_{n}}\mu_{v}(q,n)\ln \mu_{v}(q,n)
\end{equation*}
\begin{equation*}
    = \lim_{v\to 0}\frac{1}{\ln n}\sum_{v=1}^{N_{n}}\mu_{v}(q,n)\ln \mu_{v}(q,n). \tag{6}\label{eq: 6}
\end{equation*}

For values of $q$ yielding a strong relationship between Eqs.~(\ref{eq: 5}) \& (\ref{eq: 6})---in this study, correlation coefficient $r>0.95$, the parametric curve $(\alpha(q),f(q))$ or $(\alpha,f(\alpha))$ constitutes the multifractal spectrum and $\Delta\alpha$ (i.e., $\alpha_\mathrm{max}-\alpha_\mathrm{min}$) constitutes the multifractal spectrum width. $r$ determines that only scaling relationships of comparable strength can support the estimation of the multifractal spectrum, whether generated as cascades or surrogates. Using a correlation benchmark aims to operationalize previously raised concerns about mis-specifications of the multifractal spectrum \cite{zamir2003critique}.

\subsubsection{Calculating {\boldmath $t_\mathrm{MF}$} based on comparison with Iterated Amplitude Adjusted Fourier Transform (IAAFT) surrogates}

Surrogate testing using Iterated Amplitude Adjusted Fourier Transformation (IAAFT) generated $\textit{t}_\mathrm{MF}$. To identify whether nonzero $\Delta\alpha$ reflected multifractality due to nonlinear interactions across timescales, $\Delta\alpha$ for the original and $32$ IAAFT surrogates \cite{ihlen2012introduction,schreiber1996improved}. IAAFT randomizes original values time-symmetrically around the autoregressive structure, generating surrogates that randomized phase ordering of the series’ spectral amplitudes while preserving linear temporal correlations. The one-sample \textit{t}-statistic (henceforth, $\textit{t}_\mathrm{MF}$) takes the subtractive difference between $\Delta\alpha$ for the original series and that for the $32$ surrogates, dividing by the standard error of the spectrum width for the surrogates.

\subsection{Estimating ergodicity breaking parameter for cascade series}

The dimensionless statistic $EB$, the Thirumalai-Mountain metric can quantify the degree to which a series breaks ergodicity \cite{he2008random,thirumalai1989ergodic,rytov1989principles} as the variance of sample variance divided by the total-sample squared variance:
\begin{equation*}
    EB(x(t))=\frac{\Bigl \langle \Bigl[\overline{\delta^{2}(x(t))}\Bigl]^{2}\Bigl \rangle-\Bigl \langle \overline{\delta^{2}(x(t))}\Bigl \rangle^{2}}{\Bigl \langle \overline{\delta^{2}(x(t))} \Bigl \rangle^{2}}, \tag{7}\label{eq: 7}
\end{equation*}
where $\overline{\delta^{2}(x(t))}=\int_{0}^{t-\Delta}[x(t^{\prime}+\Delta)- x(t^{\prime})]^{2}dt^{\prime}\bigl/(t-\Delta)$ is the time average mean-squared displacement of the stochastic series $x(t)$ for lag time $\Delta$. Ergodicity with least breaking appears as rapid decay of $EB$ to $0$ for progressively larger samples, that is, $EB\rightarrow0$ as $t\rightarrow \infty$. Thus, for Brownian motion $EB(x(t))=\frac{4}{3}(\frac{\Delta}{t})$ \cite{cherstvy2013anomalous,metzler2014anomalous}. Slower decay indicates progressively more ergodicity breaking as in systems with less reproducible or representative trajectories, and no decay or convergence to a finite asymptotic value indicates strong ergodicity breaking \cite{deng2009ergodic}. $EB(x(t))$ thus allows estimating the extent to which a given time series fulfills ergodic assumptions or breaks ergodicity. Despite the traditional convention of respecting ergodicity as a dichotomy, $EB$ offers a window on how continuous processes can exhibit gradually more or less breaking of ergodicity \cite{kelty2022fractal,kelty2023multifractaldescriptors,mangalam2021point,mangalam2022ergodic,mangalam2023ergodic}. For instance, Deng and Barkai \cite{deng2009ergodic} have shown that for $fBm$,
\begin{multline*}
    EB(x(t))=\\ 
    \begin{cases}
        k(H_\mathrm{fGn})\frac{\Delta}{t}&0<H_\mathrm{fGn}<\frac{3}{4}\\
        k(H_\mathrm{fGn})\frac{\Delta}{t}\ln{t}&H_\mathrm{fGn}=\frac{3}{4}\\
        k(H_\mathrm{fGn})(\frac{\Delta}{t})^{4-4H_\mathrm{fGn}}&\frac{3}{4}<H_\mathrm{fGn}<1.
    \end{cases}
    \tag{8}\label{eq: 8}
\end{multline*}

Just like in multifractal analysis, estimates of ergodicity-breaking can exhibit variability based on the sequence shape or histogram. Since we are assessing the potential of $t_\mathrm{MF}$ to discern nonlinearity, our primary focus lies in sequence-specific ergodicity-breaking as a potential contributor to the variance in $t_\mathrm{MF}$. Although the raw $EB$ values may differ between different original series, we compare the $EB$ for the original series with that of shuffled versions of the same series. To achieve this, we employ the IAAFT method in the multifractal surrogate test to disrupt nonlinear temporal sequences while preserving linear ones. Considering that the IAAFT method already retains linear structures for both the multifractal test and the computation of $t_\mathrm{MF}$, we resort to shuffling exclusively to disrupt all sequential structures, whether linear or nonlinear. It is plausible that any portion of the ergodicity-breaking sequence may be a more effective predictor of $t_\mathrm{MF}$ than the multiplicative interactions across scales. Consequently, we calculate $EB$ values for both the original cascades and their shuffled counterparts, spanning generations $9$ through $15$ (with a range of $T/8$ and a lag of $\Delta=2$ samples). In our subsequent analysis, our focus shifts away from the raw magnitude of $EB$ to center on the divergence of $EB$ across timescales between the original and shuffled series. This shift in focus allows us to address the sequence-related aspects that previous applications of $t_\mathrm{MF}$ have aimed to explore \cite{kelty2022fractal,kelty2023multifractaldescriptors,mangalam2022ergodic,mangalam2023ergodic}.

\section{Results}

\begin{figure*}
    \includegraphics[width=6.75in]{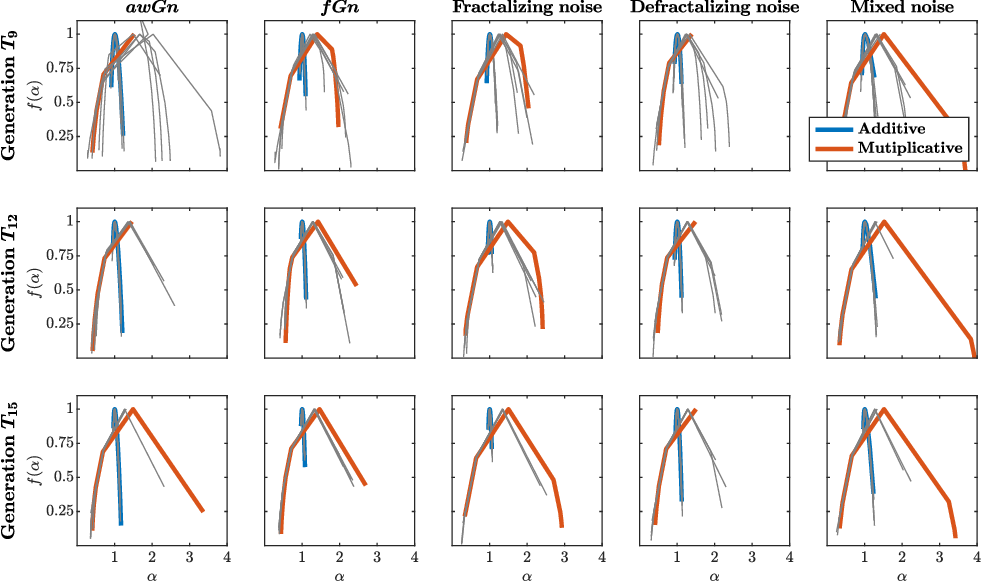}
    \caption{Multifractal spectra for representative series of the five types of random additive and multiplicative cascades across successive generations. Colored lines depict the original spectra, whereas grey lines depict five representative IAAFT surrogate spectra for each original spectrum.}
    \label{fig: f2}
\end{figure*}

\begin{figure*}
    \includegraphics[width=6.5in]{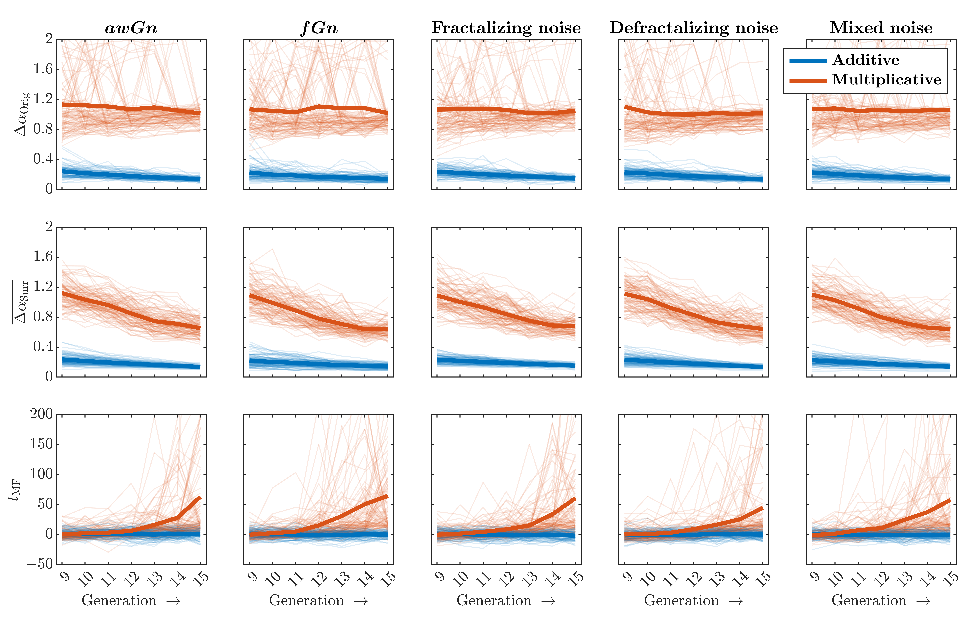}
    \caption{Multifractal spectral properties of the five types of random additive and multiplicative cascades across successive generations. Multifractal spectrum width for the original cascades $\Delta\alpha_\mathrm{Orig}$ (\textit{top}), multifractal spectrum width for the corresponding $32$ IAAFT surrogates $\overline{\Delta\alpha_\mathrm{Surr}}$ (\textit{middle}), and multifractal nonlinearity $t_\mathrm{MF}$ (\textit{bottom}). Notably, the series length $l$ experiences exponential growth across these generations: specifically, $l=2^{8},2^{9},2^{10},2^{11},2^{12},2^{13},\;\mathrm{and}\;2^{14}$ for the $9$th through $15$th generation, respectively. The depicted thick lines encapsulate mean values across $N=100$ numerical simulations, while the accompanying thin lines capture individual instances within these simulations.}
    \label{fig: f3}
\end{figure*}

\subsection{Hypothesis 1: Random multiplicative cascades were more likely than random additive cascades to have wider multifractal spectra with greater generations}

For the first two hypotheses, we examined how the multifractal properties---multifractality and multifractal nonlinearity---grow across successive generations of random additive and multiplicative cascades. Initial visual inspection provides an impression of how these multifractal spectra differ across the different types of cascade simulations. Fig.~\ref{fig: f2} shows an example of a multifractal spectrum for the additive and multiplicative variants of cascades applying each of the five noise types at three generations. Modeling the variance in the multifractal spectrum $f(\alpha(q)),\alpha(q)$ elaborates this case, making two important points: first, confirming that repeated generations of multiplicative interactions across scale yields wider multifractal spectra and, second, revealing that $fGn$ accentuated these differences. Using the function \texttt{lmer} from the package \texttt{lme4} \cite{bates2009package} in $R$ \cite{team2013r}, we ran a linear mixed-effect regression model of three different multifractal-spectral attributes to elaborate beyond these cursory portrayals. Our regression model included four covariates:
\begin{enumerate}
    \item \textbf{\textit{Feature}} encoding three types of multifractal-analytical outcomes \cite{morales2002wavelet}, namely, a baseline value of $\alpha_\mathrm{max}-\alpha(q=0)$ (i.e., spectral half-width to the \textit{right} of the peak) and two alternative values of $\alpha(q=0)-\alpha_{min}(q)$ (spectral half-width to the \textit{left} of the peak) and $\alpha(q=0)$ (i.e., the location of the spectral peak). 
    \item \textbf{\textit{Type}} encoding the five types of cascade as described above, with a baseline value of $awGn$ and four alternative values $fGn$, \textit{fractalizing}, \textit{defractalizing}, and \textit{mixed}. 
    \item \textbf{\textit{Multiplicativity}} encoding the way that cascade simulations applied noise across generations, with a baseline value of \textit{additive} and alternative value \textit{multiplicative}.
    \item \textbf{\textit{Generation}} encoding the number of generations after $9$ and through $15$.
\end{enumerate}
We used a full-factorial regression model, including the highest-order interaction covariate Feature $\times$ Type $\times$ Multiplicative $\times$ Generation and all component lower-order interactions and main effects. We detail the outcomes for each of the three features in the following paragraphs before detailing the outcome for each of our hypotheses.

The right-side width of the multifractal spectrum, $\alpha_\mathrm{max}-\alpha(q=0)$, tended to begin relatively large at the $9$th generation for $awGn$ and random additive cascades (Fig.~\ref{fig: f2}). The intercept term thus indicates that random additive $awGn$ cascades have a nonzero right-side width ($B=1.62\times 10^{-1}, SE=1.08\times 10^{-2}, P<0.0001$) that dwindles slowly with subsequent generations ($B=-1.03\times 10^{-2}, SE=2.94\times 10^{-3}, P<0.001$). Right-side width did not differ for random additive $fGn$ cascades from random additive $awGn$ cascades. After the initial $9$ generations, random multiplicative $awGn$ cascades had more than double the random additive $awGn$ cascade's right-side width ($B=8.58\times 10^{-2}, SE=1.53\times 10^{-2}, P<0.0001$), the random multiplicative \textit{Fractalizing} noise cascades had dramatically less right-side width ($B=-6.06\times 10^{-2}, SE=2.12\times 10^{-2}, P< 0.01$), but random multiplicative $fGn$ cascades had roughly half the right-side width as with $awGn$ ($B=-4.19\times 10^{-2}, SEs=2.12\times 10^{-2}, P<0.05$, respectively). The repeated multiplicative interactions of $awGn$ across generations leads to twice more rapid diminution of the right-side width ($B=-2.26\times 10^{-2}, SE=4.16\times 10^{-3}, P<0.0001$) than in the additive case, but random multiplicative \textit{mixed} noise and $fGn$ cascades shows less narrowing ($B=1.44\times 10^{-2}, SE=5.88\times 10^{-3}, P<0.05$) and none of the narrowing ($B=2.31\times 10^{-2}, SE=5.88\times 10^{-3}, P<0.0001$), respectively, with successive generations. Random multiplicative \textit{defractalizing} noise cascades showed no differences from the random additive $awGn$ cascades ($p>0.05$). Because the right-side width of the multifractal spectrum refers to the heterogeneity of small-sized fluctuations, these results mean that multiplicative interactions across scales contribute to the heterogeneity of small-sized fluctuations. Repeated multiplicative interactions of $awGn$ destroy such heterogeneity across generations, and those of unadulterated $fGn$ and $fGn$-and-$awGn$ mixtures preserve this small-sized heterogeneity progressively more than repeated multiplicative interactions of $awGn$. Hence, the effect of multiplicativity was to initially produce wider right sides of multifractal spectra (i.e., by the 9th generation) and then to slow (with \textit{mixed} noise) and cancel (with $fGn$) and the reduction in right-side width that random additive $awGn$ and random additive $fGn$ cascades showed. Perhaps the homogeneity of $fGn$ across time was sufficient to maintain comparable heterogeneity for small events.

The left-side width of the multifractal spectrum, $\alpha(q=0)-\alpha_{min}(q)$ was much wider in the random multiplicative cascades with $awGn$ ($B=7.31\times 10^{-1}, SE=2.12\times 10^{-2}, P<0.0001$) and much narrower in the random additive cascades with $awGn$ ($B=-8.52\times 10^{-2}, SE=1.50\times 10^{-2}, P<0.0001$). This increase in width on the left side is the biggest effect on both sides of the width, outstripping by more than an order of magnitude even the largest decreases on the right-side width (e.g., for multiplicative interactions repeated multiplicative interactions of $fGn$ and of \textit{Fractalizing} noise, as well as for random multiplicative $awGn$ cascades). However, while successive multiplicative generations of $awGn$ led to the progressively wider left side of the spectrum ($B=4.45\times 10^{-2}, SE=5.88\times 10^{-3}, P<0.0001$), the multiplicative interactions in $fGn$ cascades showed much slower growth of the left-side width across generations ($B=-3.31\times 10^{-2}, SE=8.31\times 10^{-3}, P<0.0001$). Random multiplicative cascades with \textit{fractalizing}, \textit{defractalizing}, and \textit{mixed} noises showed no significant differences from the random additive cascades with $awGn$ ($p\mathrm{s}>0.05$). Because the left-side width of the multifractal spectrum refers to the heterogeneity of large-sized fluctuations, these results imply that multiplicative interactions across scales, especially when involving $awGn$, contribute to the heterogeneity of large-sized fluctuations. Repeated multiplicative interactions of $awGn$ and, to a lesser extent, $fGn$ appear to accentuate this heterogeneity across generations. Hence, the effect of multiplicativity serves to accentuate drastically wider multifractal spectra on the left side to the degree that ensures random multiplicative cascades generally have greater multifractal spectrum widths $\Delta\alpha_\mathrm{Orig}$ than random additive cascades do. The only qualification of this finding is that random multiplicative $fGn$ cascades appear to slow this widening on the left side of the spectrum---although the $fGn$ cascades stabilized right-side width, it could be that the homogeneity of $fGn$ across time was insufficient to stop the growth of heterogeneity for large events but still managed to slow the growth of this large-scale heterogeneity.

These results regarding the right- and left-side width confirm Hypothesis 1, the expectation that random multiplicative cascades would have generally wider multifractal spectrum widths. This point may have been apparent from the visual inspection of Fig.~\ref{fig: f2}. What the regression modeling shows that visual inspection may not be is that applying $fGn$ towards the multiplicative interactions across scales might dampen the sensitivity of multifractal spectrum width to successive generations.

Lastly, comparably to the left-side width $\alpha(q=0)-\alpha_{min}(q)$ (Fig.~\ref{fig: f2}), the horizontal location of the multifractal spectrum peak---encoded by $\alpha(q=0)$---increased with $fGn$, multiplicativity, and repeated multiplicative interactions of $awGn$ across generations (Fig.~\ref{fig: f2}). Random multiplicative cascades with $awGn$ had greater $\alpha(q=0)$ ($B=3.91\times 10^{-1}, SE=2.12\times 10^{-2}, P<0.0001$) than the nonzero but smaller $\alpha(q=0)$ for random additive $awGn$ cascades ($B=8.48\times 10^{-1}, SE=1.50\times 10^{-2}, P<0.0001$). Successive generations of the random additive cascades with $awGn$ led to progressively larger  $\alpha(q=0)$ ($B=9.24\times 10^{-3}, SE=4.16\times 10^{-3}, P<0.05$), but the increase in location $\alpha(q=0)$ was almost three times larger with successive generation of the random multiplicative cascade applying $awGn$ ($B=2.66\times 10^{-2}, SE=5.88\times 10^{-3}, P<0.001$). Multiplicative interactions in $fGn$ cascades canceled out such increase in $\alpha(q=0)$ across generations ($B=-2.78\times10^{-2}, SE=8.31\times10^{-3}, P< 0.001$). \textit{Fractalizing}, \textit{defractalizing}, and \textit{mixed} cascades showed no differences from the random additive $awGn$ cascades. The horizontal location of the multifractal spectrum peak encodes the scale dependence of bin proportions with equal weighting of all bins, and more specifically, it is the singularity strength calculated without the presumption of heterogeneity of large or small events. Thus, these results mean that multiplicative interactions across scales and $fGn$ are major contributors to the singularity of these series if we consider them without the heterogeneity-accentuating lens of the $q$ exponent. Whereas repeated generations of $awGn$ had reduced the heterogeneity of small-size fluctuations indicated by right-side width, it appears that repeated generations of multiplicative interactions of $awGn$ did increase the homogeneous singularity strength of the cascade series.

\subsection{Hypothesis 2A: Progressively more generations of random multiplicative cascades increased multifractal nonlinearity by reducing surrogates' multifractal spectrum width}

Because Chhabra \& Jensen’s \cite{chhabra1989direct} method for estimating multifractal spectrum requires a minimum length of $l\sim2^8$ to have at least four separate scales in the computation of the power-law relationship corresponding to each $q$ exponent, we analyzed data from $9$th generation onward until the $15$th generation, yielding cascading series of lengths $l=2^{8},2^{9},2^{10},2^{11},2^{12},2^{13},2^{14}$ for $9$th, $10$th, $11$th, $12$th, $13$th, $14$th, and $15$th generations, respectively. We found that multifractality---quantified in terms of multifractal spectrum width of the original series, $\Delta\alpha_\mathrm{Orig}$---remained constant across generations (Figs.~\ref{fig: f2} \& \ref{fig: f3}, \textit{top}), but multifractal spectrum width of the corresponding IAAFT series, $\Delta\alpha_\mathrm{Surr}$ showed an asymptotic reduction across generations, approaching zero by $15$th generation (Figs.~\ref{fig: f2} \& \ref{fig: f3}, \textit{middle}). Consequently, multifractal nonlinearity---quantified in terms of $t$-statistics comparing $\Delta\alpha_\mathrm{Orig}$ and $\Delta\alpha_\mathrm{Surr}$, $t_\mathrm{MF}$---showed an exponential increase across generations, approaching $\sim 750$ by $15$th generation (Fig.~\ref{fig: f3}, \textit{bottom}).

The distinction between the original and surrogate spectrum widths suggests that, with subsequent generations of parents and children, the original series' multifractality departs more and more from the multifractality attributable to the linear structure of IAAFT surrogates, accentuating the need for a nonlinear model for the analysis of cause and effect. Notably, the further generations do not appear to increase the estimated multifractal spectrum width for the original series (Figs.~\ref{fig: f2} \& ~\ref{fig: f3}, \textit{top}). Multifractal spectrum width is relatively constant across generations as well. Instead, the multifractal spectrum width for surrogates appears to narrow with a greater number of generations (Figs.~\ref{fig: f2} \& ~\ref{fig: f3}, \textit{middle}). Hence, visual inspection suggests that random multiplicative cascades yielded larger $\Delta\alpha_\mathrm{Orig}$, smaller $\Delta\alpha_\mathrm{Surr}$, and larger $t_\mathrm{MF}$ than random additive cascades (Fig.~\ref{fig: f3}). This pattern appeared robust across all noise types, and indeed, again, from visual inspection of these plots, it seems that the only observable change was the decrease in $\Delta\alpha_\mathrm{Surr}$ and increase in $t_\mathrm{MF}$ with greater numbers of generations and length. In summary, random multiplicative cascades exhibit larger $t_\mathrm{MF}$ than random additive cascades. Our randomization of these multiplicative cascades using $awGn$ appears to have led successive generations to small but systematic decreases in $t_\mathrm{MF}$. Nonetheless, multiplicativity positively affected $t_\mathrm{MF}$, suggesting a slowing of this decrease with generation.

\subsubsection{Repeated generations of multiplicative interactions across scales increased the continuous value of $t_\mathrm{MF}$, especially for random multiplicative $fGn$ cascades}

To test these observations of continuous variation more formally, we used the function \texttt{lmer} from the package \texttt{lme4} \cite{bates2009package} in $R$ \cite{team2013r} to test an analogous regression model as used for Hypothesis 1. Instead of testing the features of the original multifractal spectra, we now tested these three different multifractal outcomes, that is, $\Delta\alpha_\mathrm{Orig}$ $\Delta\alpha_\mathrm{Surr}$, and $t_\mathrm{MF}$. Whereas the regression model supporting Hypothesis 1 used a predictor \textit{Feature}, we now defined the analogous covariate \textit{Outcome} to have baseline value of  $\Delta\alpha_\mathrm{Orig}$ and alternate values $\Delta\alpha_\mathrm{Surr}$ and $t_\mathrm{MF}$. We used the linear mixed-effect regression model testing the higher-order interaction Outcome $\times$ Type $\times$ Multiplicative $\times$ Generation and all component lower-order interactions and main effects. The significant effects of this regression modeling were remarkably few, numbering only $3$, suggesting that the effects of multiplicativity and noise type are stronger on $t_\mathrm{MF}$ than on $\Delta\alpha_\mathrm{Orig}$ or $\Delta\alpha_\mathrm{Surr}$. First, at the $9$th generation, random multiplicative cascades actually exhibited lower $t_\mathrm{MF}$ than additive cascades ($B=-1.24\times 10^{1}, SE=3.02\times 10^{0}, P<0.0001$). However, second, repeated generations specifically of multiplicative interactions across time scale increase $t_\mathrm{MF}$ ($B=9.08\times 10^{0}, SE=8.37\times 10^{-1}, P<0.0001$), suggesting that $t_\mathrm{MF}$ becomes positive on the $10$th generation and continues to increase steadily with subsequent generations. This increase in $t_\mathrm{MF}$ with subsequent generations of multiplicative interactions is even faster for random multiplicative cascades using $fGn$ ($B=2.37\times 10^{0}, SE=1.18\times 10^{0}, P<0.05$), indicating that $t_\mathrm{MF}$ for random multiplicative $fGn$ cascades might be closer to zero than negative on the 9th generation. What is curious about these findings overall is that, in support of Hypothesis 1, we have found that $fGn$ slowed any change to multifractal-spectrum width across generations. But now we see that, within that narrower multifractal spectrum, enacting the random multiplicative cascade with $fGn$ may produce greater signatures of nonlinear correlations across scales than cascades with less correlated noise patterns.

\subsubsection{Repeated generations of multiplicative interactions across scale increased the frequency of significant positive values of $t_\mathrm{MF}$, especially for random multiplicative $fGn$ cascades}

We modeled the $7$ seven generations from the $9$th to $15$th generations of the $1000$ cascade processes across additive and multiplicative cascades with five kinds of noises, yielding $7000$ individual generations of random cascade processes. Out of these $7000$, there were $3592$ significant positive values of $t_\mathrm{MF}$ indicating significantly wider multifractal spectra for the original series compared to a sample of surrogate series' multifractal spectra for 51.31\% of the simulated generations. 

To test the observations of dichotomous multifractal nonlinearity, that is, the number of significant positive values of $t_\mathrm{MF}>2.04$, we used a mixed-effect logistic regression with the function \texttt{glmer} from the package \texttt{lme4} \cite{bates2009package} in $R$ \cite{team2013r}. We attempted to use the same family of higher-order and component lower-order interactions and main effects, but these models did not converge. Therefore, we trimmed the model to test only the sum of two interactions, namely, Type $\times$ Multiplicative and Type $\times$ Generation, as well as all component lower-order interactions and main effects. The results here followed the same directions as the significant effects on continuous variation of $t_\mathrm{MF}$. We then found additional effects of different types of noise that the model of continuous variation of $t_\mathrm{MF}$ did not show. Much like the model of continuous variation of $t_\mathrm{MF}$, the results of the logistic model testing for differences in frequency of significant resulting model showed a negative effect of random multiplicative $awGn$ cascades ($B=-1.50\times 10^{0}, SE=1.59\times 10^{-1}, P<0.0001$) as well as a positive effect of generations for random multiplicative $awGn$ cascades ($B=7.66\times 10^{-1}, SE=3.13\times 10^{-2}, P<0.0001$). These effects suggest that the odds ratio for significant positive $t_\mathrm{MF}>2.04$ begins negatively on the $9$th generation, goes to zero on the $11$th generation, and becomes positive on the $12$th generation. Interpreting this change in odds ratio, we can say that this model suggests that the significant positive $t_\mathrm{MF}$ are less likely in the $9$th generation but, by the $12$th generation, became more likely to occur---a finding that aligns with and adds nuance to the foregoing finding that continuous values of $t_\mathrm{MF}$ begin negative on $9$th generation and grow with further generations.

In addition, a few interaction effects in the regression model indicated that random multiplicative cascades with specific types of noise were significantly more likely to produce significant positive $t_\mathrm{MF}$. As noted above, the logistic model did not support a comparable interaction effect for the effect of subsequent generations in specifically random multiplicative $fGn$ cascades. However, in a similar vein as the regression model for continuous variation in $t_\mathrm{MF}$ had shown a positive effect of progressive generations in specifically random multiplicative $fGn$ cascade, this logistic regression indicated the random multiplicative $fGn$ cascades had higher likelihood of significant positive $t_\mathrm{MF}$ irrespective of generation ($B=4.30\times 10^{-1}, SE=1.74\times 10^{-1}, P<0.05$). Also irrespective of generation, random additive \textit{fractalizing} noise cascades and random additive \textit{mixed} noise cascades had lower likelihood of significant positive $t_\mathrm{MF}$ ($B\mathrm{s}=-2.26\times10^{-1}\;\mathrm{and}\;-2.39\times 10^{-1}, SE=1.11\times10^{-1}\;\mathrm{and}\;1.11\times 10^{-1}, P<0.05$, respectively), and random multiplicative \textit{fractalizing} and random multiplicative \textit{mixed} noise cascades had higher likelihood of significant positive $t_\mathrm{MF}$ ($Bs=4.68\times 10^{-1} and 3.64\times 10^{-1}, SE=1.74\times 10^{-1} and 1.73\times 10^{-1}, P<0.05$, respectively). Hence, multiplicative interactions that apply noise with greater contributions of $fGn$ appear to show a higher frequency of significant positive $t_\mathrm{MF}$---or multifractal nonlinearity.

\subsubsection{Repeated generations of multiplicative interactions across scale \textit{decreased} the frequency of significant \textit{negative} values of $t_\mathrm{MF}$, especially for random multiplicative $fGn$ cascades}

As further evidence that random multiplicative cascades contribute to the greater frequency of significantly positive $t_\mathrm{MF}$ (i.e., $t_\mathrm{MF}>2.04$), we can demonstrate the opposite, namely that repeated generation of additive cascades can increase the likelihood of significant \textit{negative} $t<-2.04$. Within our sample of $7000$ simulated generations, we found $1895$ such significant negative $t_\mathrm{MF}$, $27.07\%$ of the total simulated generations. Negative differences between multifractal spectrum width for original and surrogates receive less attention in the theoretical literature \cite{lee2017cascade}. However, the empirical evidence has regularly found these negative $t_\mathrm{MF}$ to be a persistent minority of the multifractal results. Notably, these negative $t_\mathrm{MF}$ have regularly worked alongside positive $t_\mathrm{MF}$ to serve as one continuous, real-numbered covariate useful for predicting behavioral and biological outcomes \cite{bell2019non,carver2017multifractal,carver2017multifractality,doyon2019multifractality,harrison2014multiplicative,jacobson2021multifractality,kelty2018multifractal,kelty2021multifractal,kelty2023multifractalnonlinearity,mangalam2020multifractal,mangalam2020multiplicative,palatinus2014haptic,stephen2012multifractal}. 

We saw the opportunity to use regression modeling to understand how these negative $t_\mathrm{MF}$ cases compare to positive $t_\mathrm{MF}$ regarding their sensitivity to multiplicative interactions across scales. In this negative case, the smaller set of values entailed that a logistic regression enlisting the same covariates did not converge. So, we trimmed the model to the nearest subset of those covariates that did converge, namely the sum of the main effect Type and the interaction effect Multiplicative $\times$ Generation, as well as both constituent main effects of the latter interaction. 

The general pattern of findings showed that negative $t_\mathrm{MF}$ appeared to result from fewer generations of multiplicative noise. Negative $t_\mathrm{MF}$ resulted from random multiplicative cascades sooner (i.e., with fewer generations) than random additive cascades. The random additive cascades were less likely to yield significant negative $t_\mathrm{MF}$ using $awGn$ ($B=-9.19\times 10^{-1}, SE=1.02\times 10^{-1}, P<0.0001$), but this likelihood increased over subsequent generations of random additive $awGn$ cascades ($B=4.51\times 10^{-2}, SE=9.29\times 10^{-2}, P<0.05$). The role of multiplicativity in contributing to the likelihood of negative $t_\mathrm{MF}$ mirrored the directions for the effects of multiplicativity on the likelihood of positive $t_\mathrm{MF}$. Specifically, random multiplicative cascades initially (i.e., at the $9$th generation) showed a significantly greater likelihood of yielding negative $t_\mathrm{MF}$ ($B=1.01\times 10^{0}, SE=1.20\times 10^{-1}, P<0.0001$). Nonetheless, successive generations of random multiplicative cascades predicted a dramatically lower likelihood of negative $t_\mathrm{MF}$ ($B=-8.15\times 10^{-1}, SE=3.83\times 10^{-2}, P<0.0001$). Notably, the latter effect is almost $20$ times stronger and in the opposite direction of the effect of gradual increase in the likelihood of negative $t_\mathrm{MF}$ with more generations of the random additive $awGn$ cascades.

Finally, we found a significant increase in the likelihood of negative $t_\mathrm{MF}$ for random additive \textit{defractalizing} cascades, suggesting that applying noise that is progressively less correlated can produce significantly negative $t_\mathrm{MF}$. This point may have little to do with the effect of multiplicativity and more to do with previous observations that increasing the strength of $awGn$ noise terms can lead cascade dynamics to yield progressively more negative $t_\mathrm{MF}$ \cite{lee2017cascade}.

\subsection{Hypothesis 2B: Length of nonoverlapping segments and length across overlapping segments both increased multifractal nonlinearity by reducing surrogates' multifractal spectrum width}

\begin{figure*}
    \includegraphics[width=6.5in]{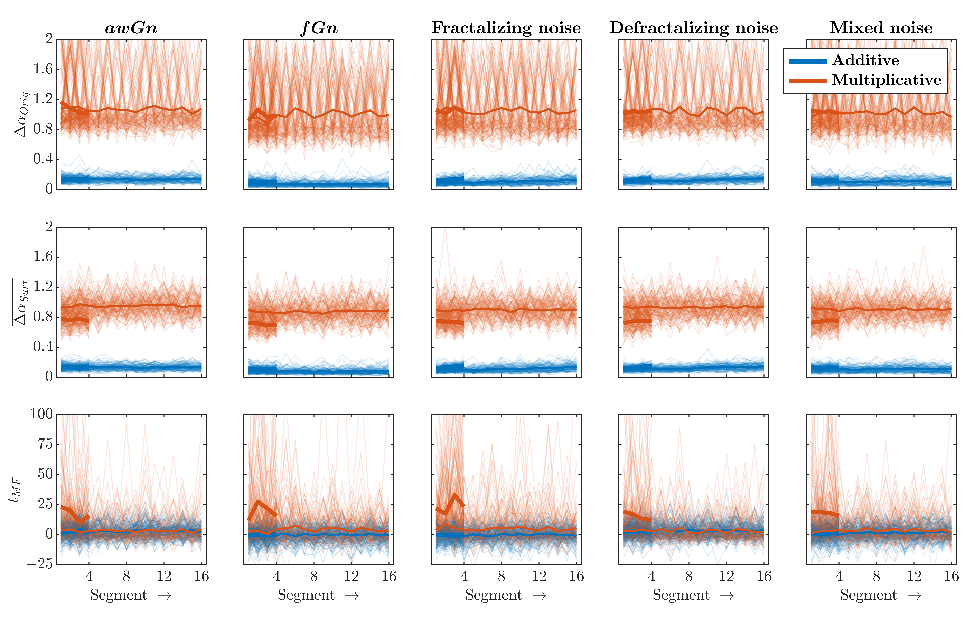}
    \caption{Multifractal spectral properties of the five random additive and multiplicative cascade types across non-overlapping segments in $15$th and final generation. Multifractal spectrum width for the original cascades $\Delta\alpha_\mathrm{Orig}$ (\textit{top}), multifractal spectrum width for the corresponding $32$ IAAFT surrogates $\overline{\Delta\alpha_\mathrm{Surr}}$ (\textit{middle}), and multifractal nonlinearity $t_\mathrm{MF}$ (\textit{bottom}). Thick and thin lines indicate segments of length $l=2^{12}\;\mathrm{and}\;2^{10}$, respectively. Each line encapsulates mean values across $N=100$ numerical simulations.}
    \label{fig: f4}
\end{figure*}

The preliminary view on Hypothesis 2A is that $t_\mathrm{MF}$ sooner reflects a stable $\Delta\alpha_\mathrm{Orig}$ and a dwindling $\Delta\alpha_\mathrm{Surr}$. This finding is reassuring on the point that progressively more generations of a random multiplicative process might yield a larger $t_\mathrm{MF}$. However, this conclusion warrants further investigation because simply iterating progressively more generations of a cascade process by replacing a single ``parent'' value in an earlier generation makes a longer series. So, length is confounded with generation number when we necessarily replace fewer parents with more children from one generation to another.

The multifractal nonlinearity appears to be influenced by the length factor, potentially due to narrowing the surrogate spectrum. This observation challenges the notion that $t_\mathrm{MF}$ is exclusively sensitive to repeated multiplicative interactions across scales and suggests that a longer phase-randomized series has a greater capacity to exhibit enhanced homogeneity in temporal structure compared to the original cascade series. However, before delving into a comprehensive analysis of the length-generation interplay, it is essential to consider that the length effect may stem from statistical artifacts inherent to cascade dynamics. Addressing these artifacts is a prerequisite to establishing the genuine length dependence of $t_\mathrm{MF}$. For instance, subjecting the outcome of the cascade model to multifractal analysis in its entirety may overly emphasize certain idiosyncrasies of cascade models. One of these idiosyncrasies is the well-documented intermittency of cascade series, characterized by the systematic proliferation of increasingly larger events with diminishing probability as their size increases\cite{bak1987self,bak1988self,lovejoy2018weather,mandelbrot1974intermittent}. This phenomenon might exert a stronger influence on progressively longer series. To explore this, we investigated the length dependence within longer cascade series by dividing them into nonoverlapping segments of varying lengths, including $l=2^{8},2^{9},2^{10},2^{11},2^{12},2^{13}$, resulting in $2^{6},2^{5},2^{4},2^{3},2^{2},2^{1}$ segments of each length, respectively, with the last segment spanning the entire series. Remarkably, this analysis revealed consistent length dependencies across these nonoverlapping segments within the $15$th generation of the cascade model, including stable $\Delta\alpha_\mathrm{Orig}(q)$, declining $\Delta\alpha_\mathrm{Surr}(q)$, and increasing $t_\mathrm{MF}$ (Fig.~\ref{fig: f4}).

\begin{figure*}
    \includegraphics[width=6.5in]{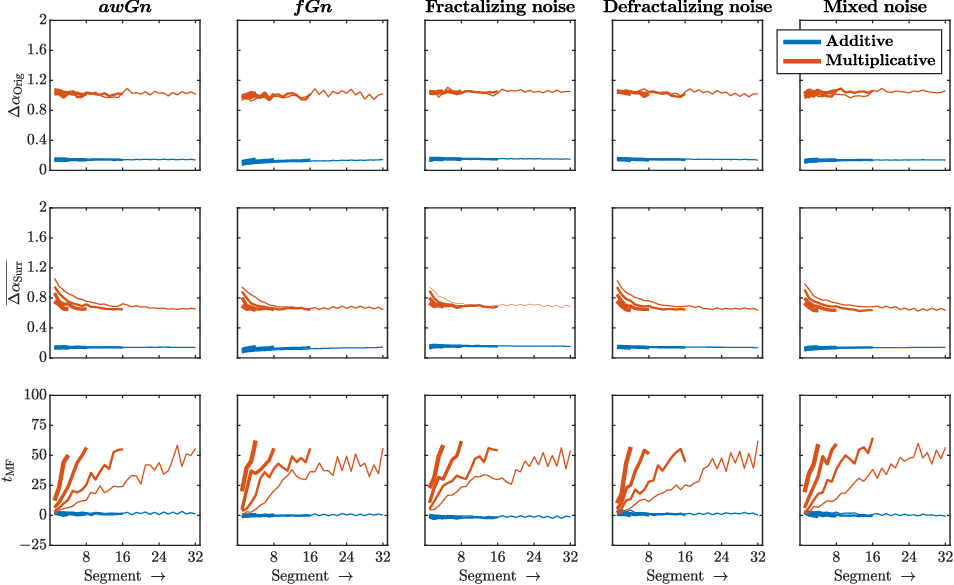}
    \caption{Multifractal spectral properties of the five types of random additive and multiplicative cascades across progressively longer segments from the beginning to the end in $15$th and final generation. Multifractal spectrum width for the original cascades $\Delta\alpha_\mathrm{Orig}$ (\textit{top}), multifractal spectrum width for the corresponding $32$ IAAFT surrogates $\overline{\Delta\alpha_\mathrm{Surr}}$ (\textit{middle}), and multifractal nonlinearity $t_\mathrm{MF}$ (\textit{bottom}). The segment length $l$ reduces from $2^{12}$ to $2^{11}$, $2^{10}$ to $2^{9}$ from the thickest to the thinnest lines. Each line encapsulates mean values across $N=100$ numerical simulations.}
    \label{fig: f5}
\end{figure*}

\begin{figure*}
    \includegraphics[width=6.5in]{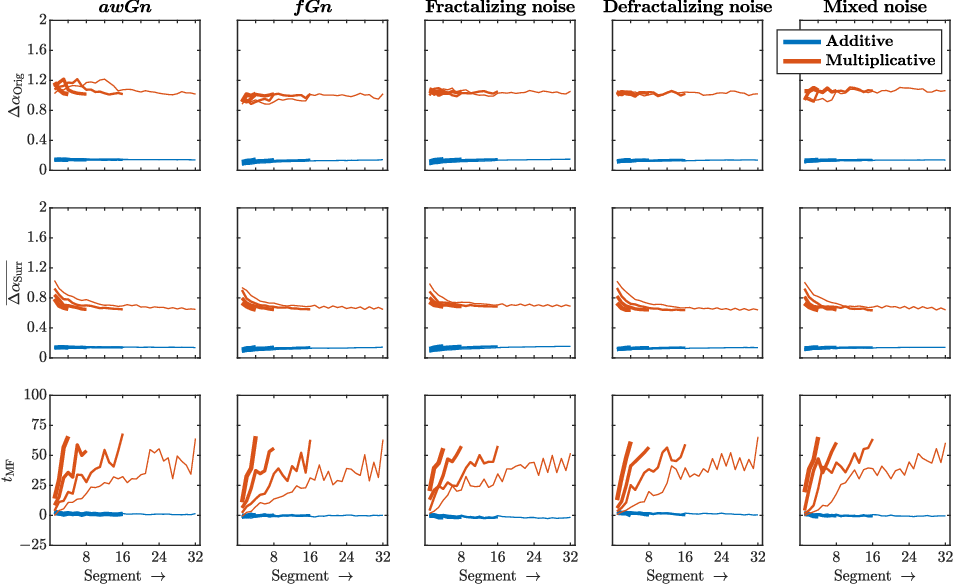}
    \caption{Multifractal spectral properties of the five types of random additive and multiplicative cascades across progressively longer segments from the end to the beginning in $15$th and final generation. Multifractal spectrum width for the original cascades $\Delta\alpha_\mathrm{Orig}$ (\textit{top}), multifractal spectrum width for the corresponding $32$ IAAFT surrogates $\overline{\Delta\alpha_\mathrm{Surr}}$ (\textit{middle}), and multifractal nonlinearity $t_\mathrm{MF}$ (\textit{bottom}). The segment length $l$ reduces from $2^{12}$ to $2^{11}$, $2^{10}$ to $2^{9}$ from the thickest to the thinnest lines. Each line encapsulates mean values across $N=100$ numerical simulations.}
    \label{fig: f6}
\end{figure*}

In a second option, related to the above but different in admitting overlapping segments, these abrupt changes across time arise from the deep ancestry of child cells from separate parents, enacting nonoverlapping windows of family relation. There emerges a potential aliasing effect: multifractal analyses using all nonoverlapping bins in Chhabra and Jensen's \cite{chhabra1989direct} method may align more clearly with the original cascade series, and randomizing the phase of these cascades series is more likely to nudge the abrupt changes from the transitions \textit{between} nonoverlapping bins of the multifractal analysis into the span of the bins. So, we tested length dependence by applying multifractal analysis to segments designed for length to overlap with the binomial fractures across even the earliest cascade generations, using natural-number multiples of our segment lengths (e.g., for $j+1$th generation of a cascade, $l=m\cdot 2^{7+s},m=1,2,3,\dots,n/2^{7+s}$, where $n$ is the series length ($=2^{14}$) and $s\leq j-7$ and anchoring, to begin with, the first value of the series $p_{1,j+1}$ (e.g., spanning $[p_{1,j+1},p_{2,j+1},\dots,p_{m\cdot 2^{7+s},j+1}]$) or to end with the last value of the series (e.g., spanning $[x_{2^{j}-m\cdot 2^{7+s}+1,j+1},x_{2^{j}-m\cdot 2^{7+s}+2,j+1},\dots,x_{2^{j},j+1}]$). We tested this option and found that these overlapping segments of growing length within a given cascade showed the same length dependence: stable $\Delta\alpha_\mathrm{Orig}(q)$, decreasing $\Delta\alpha_\mathrm{Surr}$, and increasing $t_\mathrm{MF}$ for longer nonoverlapping segments within the $15$th and final generation of the cascade model---no matter whether progressively longer segments were drawn from the beginning and grew towards the end of the $15$th generation of the cascade series (Fig. ~\ref{fig: f5}) or from the beginning and grew towards the end of the $15$th and final generation of the cascade series (Fig. ~\ref{fig: f6}).

These patterns suggest that $t_\mathrm{MF}$ is length-dependent. The multifractality of the original cascades departs increasingly from the multifractality attributable to the linear structure of IAAFT surrogates across progressively longer segments and so longer series. Furthermore, the fact that this departure involved a reduction in $\Delta\alpha_\mathrm{Surr}$ while $\Delta\alpha_\mathrm{Orig}$ fluctuated around the mean suggests that the observed linear increase in $t_\mathrm{MF}$ reflected a homogenization of linear correlations across longer time---and not over-representation of multifractality or nonlinear correlations across scales. Hence, it looks for now as if length dependence could well explain the growth of $t_\mathrm{MF}$ separate from any effect of repeated generations.

\subsection{Hypothesis 2C: Random multiplicative cascades accrue greater multifractal nonlinearity $t_\mathrm{MF}$ when controlling for length}

A major interest in the above results has been the possibility that the growth of $t_\mathrm{MF}$ beyond zero reflects nonlinear interactions across scales due to repeated generations of multiplicative cascade dynamics. Although the results in favor of Hypothesis 2A point favorably in this direction, the results favoring Hypothesis 2B warn against a less interesting finding: greater $t_\mathrm{MF}$ could result from a longer sampling of a multiplicative process. Results supporting Hypothesis 2B indicated that random multiplicative cascade series maintain the same multifractal-spectrum width $\Delta\alpha(q)_\mathrm{Orig}$ (Figs.~\ref{fig: f4}--\ref{fig: f6}). 

The ``padded'' cascades thus offer a way to determine whether the increases in $t_\mathrm{MF}$ with successive generations supporting Hypothesis 2A were not sooner explainable by length effects. The padded cascades included repetitions of values in earlier generations to match the length of the final generation. The only change across generations of padded cascades was the application of noise terms across progressively smaller fractions of the total final-generation series length. We computed the same multifractal outcomes, that is, $\Delta\alpha_\mathrm{Orig}$, $\Delta\alpha_\mathrm{Surr}$, and $t_\mathrm{MF}$. The only challenge was applying the IAAFT procedure: the padding with repeated equivalent values within the same series entailed that phase-randomized surrogates were more likely to resemble the original cascade series precisely for relatively earlier generations containing longer stretches of the same value. This tendency contributes to a zero standard deviation among surrogates. To circumvent this consequence of the padding strategy, we generated IAAFT surrogates from un-padded versions for which length was a binary power of 2, and then we applied the padding to produce IAAFT-surrogate series that would match all other series' lengths. Then, we computed multifractal analysis with the Chhabra \& Jensen technique on both the padded original series and the sample of padded IAAFT-surrogate series for subsequent calculation of $t_\mathrm{MF}$. 

\begin{figure*}
    \includegraphics[width=6.5in]{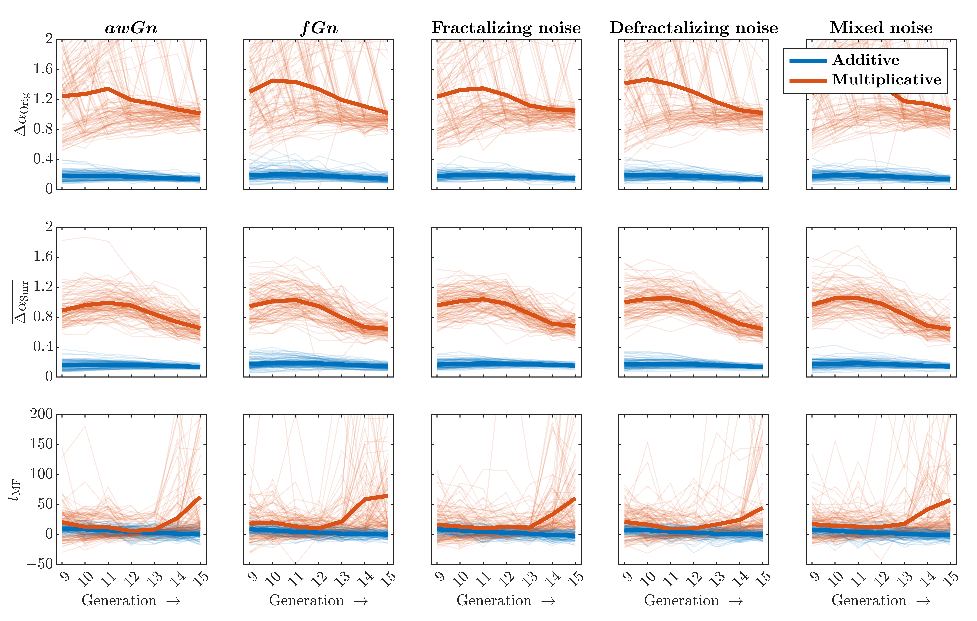}
    \caption{Multifractal spectral properties of the five types of random additive and multiplicative cascades across successive generations after controlling for the series length. Multifractal spectrum width for the original cascades $\Delta\alpha_\mathrm{Orig}$ (\textit{top}), multifractal spectrum width for the corresponding $32$ IAAFT surrogates $\overline{\Delta\alpha_\mathrm{Surr}}$ (\textit{middle}), and multifractal nonlinearity $t_\mathrm{MF}$ (\textit{bottom}). To ensure consistent length $l$ across generations, we padded the series with consecutive repetitions of individual-cell values for $9$th through $15$th generations a specific number of times: $64,32,16,8,4,2,\;\mathrm{and}\;1$ time, respectively. The depicted thick lines encapsulate mean values across $N=100$ numerical simulations, while the accompanying thin lines capture individual instances within these simulations.}
    \label{fig: f7}
\end{figure*}

When holding length constant, progressive generations of multiplicative interactions across time scale will increase multifractal nonlinearity $t_\mathrm{MF}$. As  Fig.~\ref{fig: f7} shows, progressive generations of multiplicative interactions produced gradual decreases in $\Delta\alpha_\mathrm{Orig}$ and in $\Delta\alpha_\mathrm{Surr}$ but gradual increases in $t_\mathrm{MF}$ with generation. Hence, although it is surely possible for increasing series length to increase $t_\mathrm{MF}$, it is true that the increase of $t_\mathrm{MF}$ beyond zero is also sensitive to the number of generations.

\subsection{Hypothesis 3: Random additive cascades uniformly broke ergodicity without increasing $t_\mathrm{MF}$ while ergodicity-breaking noise terms promoted ergodicity-breaking only in random multiplicative cascades}

Random additive cascades break ergodicity throughout, while random multiplicative cascades exhibit only moderate amounts of ergodicity breaking. Echoing prior remarks about our methodology, it is important to distinguish between the raw magnitude of $EB$ and the difference in the decay of $EB$ with $t$ between the original and shuffled cascades. The former differences would only refer to differences in histogram---and indeed, we may note that inevitable non-Gaussianity consequent to the multiplicative interactions within cascades \cite{castaing1990velocity,kiyono2007estimator,kiyono2008non} makes all random multiplicative cascades break more ergodicity than random additive cascades, whose repeated additions make a symmetric, short-tailed Gaussian distribution more likely. Our concern for whether $t_\mathrm{MF}$ might be specific to the original sequence of cascades requires considering the comparison to shuffled cascades to understand how the original cascades break ergodicity.

\begin{figure*}
    \includegraphics[width=6.5in]{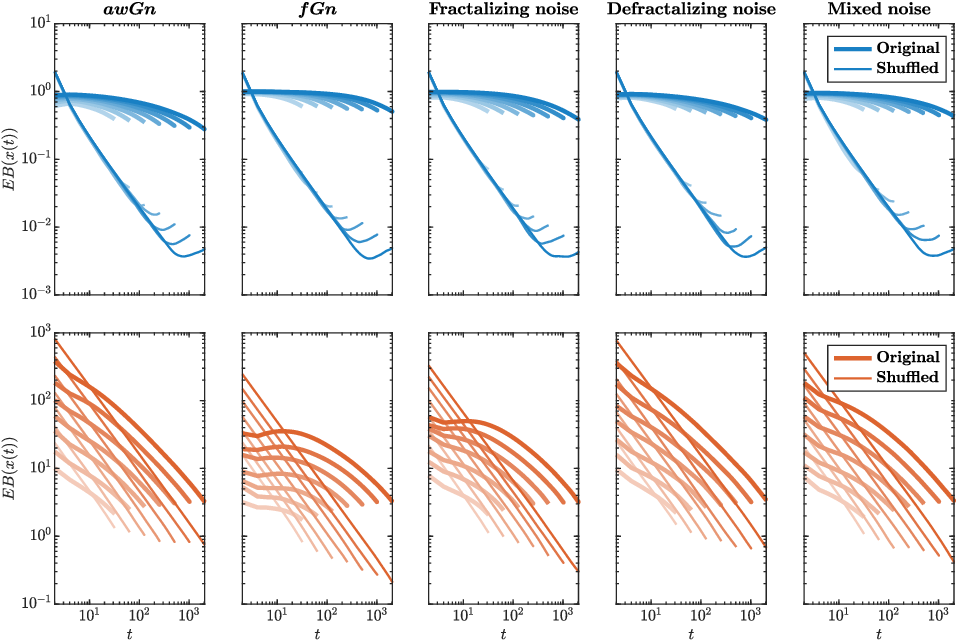}
    \caption{Ergodicity breaking in the five types of random additive (\textit{top}) and multiplicative (\textit{bottom}) cascades across successive generations ($N=100$ numerical simulations for each case, with a lag of $\Delta=10$ samples). This phenomenon is quantified using the ergodicity breaking factor $EB(x(t))$. The traces progressively deepen in color from $9$ through generation $15$. Notably, the length $l$ of the time series experiences exponential growth across these generations: specifically, $l=2^{8},2^{9},2^{10},2^{11},2^{12},2^{13},\;\mathrm{and}\;2^{14}$ for $9$th through $15$th generation, respectively.}
    \label{fig: f8}
\end{figure*}

We had hypothesized that the type of noise would influence ergodicity breaking, that is, with the degree of ergodicity breaking in the type of noise (i.e., low for $awGn$ and high for $fGn$) would amount to a comparable degree of sequence-driven ergodicity breaking in cascades, regardless of additivity or multiplicativity. In fact, as the similar pattern across all top panels of Fig.~(\ref{fig: f8}), the types of noise did not affect sequence-driven ergodicity breaking in random additive cascades. For all noise types, random additive cascades showed only very slow decay in the ergodicity-breaking statistics across scales, indicating that the variance of these processes did not converge to stable values over a long time. This pattern in the original cascades stood in contrast to the rapid decay of the ergodicity-breaking statistics for the shuffled counterparts, whose shufflings removed original temporal correlations. This disparity between the original and the shuffled cascades indicated that the ergodicity breaking was specific to the original sequence governed by the nested fracturing of the cascade process across generations---rather than to any histogram skew from a symmetric, uncorrelated Gaussian and so an inherently ergodic process.

As for random multiplicative cascades, the comparison of the original with shuffled cascades showed that sequence-dependent ergodicity breaking was markedly less for random multiplicative than for random additive cascades. Again, as noted and predicted before, the magnitude of $EB$ for random multiplicative cascades was higher than that for random additive cascades in keeping with the expectation of systematically different histograms due to lognormal versus Gaussian outcomes of multiplying versus adding. The only exception of the generally weaker sequence-driven ergodicity breaking was that random multiplicative cascades involving more generations with long-range correlated $fGn$ showed a temporarily slower decay over short to medium scales. This ergodicity breaking over short-to-medium time scales appeared in $fGn$, \textit{Fractalizing}, and \textit{Mixed} but not in the $awGn$ or the \textit{Defractalizing} cases where uncorrelated, ergodic $awGn$ dominates all generations or where the subsequent generations undo the work of correlated noise at earlier generations.

Hypothesis 3 originally posited two expectations: (i) that $fGn$ cascades would display a higher likelihood of experiencing ergodicity-breaking in their original sequences compared to cascades employing $awGn$, and (ii) that such differences in ergodicity-breaking between $fGn$-driven and $awGn$-driven cascades might yield variations in $t_\mathrm{MF}$ distinct from zero across different cascades and noise types. Our findings substantiate the first expectation but with a caveat; we observed that the noise type influenced ergodicity-breaking primarily in the case of random multiplicative cascades. Concerning the second expectation, we discovered that strong, sequence-driven ergodicity-breaking across all scales was predominantly a characteristic of random additive cascades rather than random multiplicative ones. Specifically, random additive cascades consistently exhibited robust ergodicity-breaking in their sequences, irrespective of the noise type constituting those cascades. Given the results from Hypothesis 1, which demonstrated that random additive cascades were much more inclined to exhibit zero $t_\mathrm{MF}$, the present findings imply that the presence of an ergodicity-breaking sequence does not inherently guarantee an increase in $t_\mathrm{MF}$. Our present results suggest that the nonzero $t_\mathrm{MF}$ observed in the case of random multiplicative cascades is more closely associated with weak ergodicity-breaking. This aligns with existing literature implicating multiplicative cascade dynamics in biological systems capable of inducing ergodicity-breaking behaviors \cite{bouchaud1992weak,burov2011single,kulkarni2003ergodic,li2022non,metzler2015weak}.

\begin{figure*}
    \includegraphics[width=6.5in]{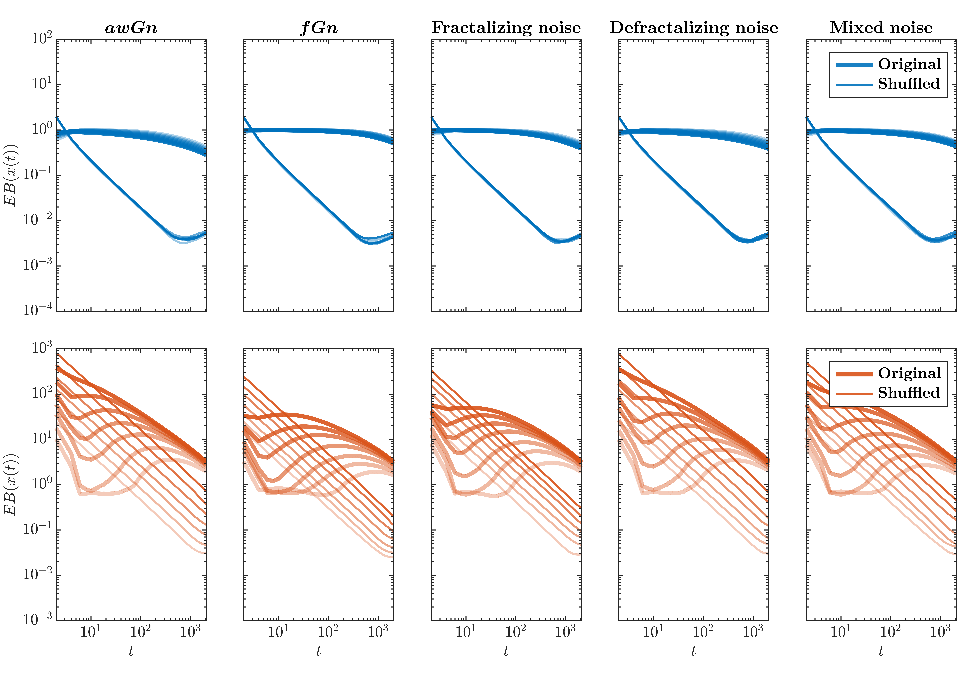}
    \caption{Ergodicity breaking in the five types of additive (\textit{top}) and multiplicative (\textit{bottom}) cascades across successive generations after controlling for the time series length ($N=100$ numerical simulations for each case, with a lag of $\Delta=10$ samples). This phenomenon is quantified using the ergodicity breaking factor $EB(x(t))$. The traces progressively deepen in color from $9$ through generation $15$. To ensure consistent length $l$ across generations, we padded the time series with consecutive repetitions of individual-cell values for $9$th through $15$th generations a specific number of times: $64,32,16,8,4,2,\;\mathrm{and}\;1$ time, respectively.}
    \label{fig: f9}
\end{figure*}

In keeping with our test of Hypothesis 2, we saw the possibility that these additivity versus multiplicativity and noise type effects could have reflected series length. Indeed, although the additive cascades showed very little variability across different generations in Fig.~\ref{fig: f7}, it is possible that the varying series lengths could have given greater expression to the ergodicity-breaking effects of $fGn$, \textit{fractalizing} noise, and \textit{mixed} noise. Hence, to ensure that the growth of sequence-driven ergodicity breaking was generation-dependent and not merely length-dependent, we submitted the same padded series for testing Hypothesis 2C. Ultimately, the only difference this controlling for length showed was in the multiplicative case, and random additive cascades showed no difference from the unpadded case (Fig.~\ref{fig: f9}). The random multiplicative cascades showed a comparable profile of ergodicity breaking with relatively more sequence-driven ergodicity breaking in the short-to-medium scales, and padding the series to control for length produced a short-range dip in $EB$ where the repeated values would have necessarily made younger generations more ergodic and to narrow the differences in ergodicity breaking across multiple generations due to noise type.

\section{Discussion}

In summary, we have shown that multifractal nonlinearity---quantified in terms of $t$-statistics comparing $\Delta\alpha_\mathrm{Orig}$ and $\Delta\alpha_\mathrm{Surr}$, $t_\mathrm{MF}$---provides a consistent, robust, and sufficient method to identify and quantify the strength of the evidence of nonlinear interactions across scales due to multiplicative cascade dynamics. We discuss how data collection and processing choices the behavioral scientist might face while performing multifractal analysis influence $t_\mathrm{MF}$ systematically and predictably. For instance, increasing measurement length may increase the observed $t_\mathrm{MF}$ for two reasons. First, longer IAAFT surrogates will have more time to exhibit the homogeneity of linear correlations compared to nonlinear correlations in the measurement series. Second, holding the length constant shows that $t_\mathrm{MF}$ is sensitive to generations. So, shorter series may be less likely to show nonzero $t_\mathrm{MF}$ because the shorter length provides less of a window on progressive generations of multiplicative interactions. Random multiplicative cascades appeared to begin producing narrower multifractal spectra than random additive cascades, and further generations served to reverse these initial differences, making random multiplicative cascades generate wider spectra than random additive cascades. The only noise constraining this expansion of random multiplicative cascades' multifractal spectra was $fGn$, and this constraint on multifractal spectra may reflect $fGn$'s homogeneous monofractality. In contrast, canonical measures like multifractal spectrum width can conflate the evidence and strength of nonlinear interactions across scales due to the differential influence of these choices on linear sources of multifractality.  

Our simulations highlight an unusual feature for new elaboration: the possibility of negative $t_\mathrm{MF}$. Specifically, this negative end of the multifractal nonlinearity has received little attention and only implicit recognition \cite{ihlen2010interaction}. Given the traditional wisdom in monofractal analysis to expect mostly greater temporal correlations as compared to shuffled surrogate series  \cite{eke2002fractal,hurst1951long,peng1994mosaic}, it may feel quite intuitive to expect only greater multifractality in the original series as compared to multifractality in the surrogate series. However, just as temporal anticorrelations can lead to shallower fractal scaling relationships than shuffled surrogates, it is important to understand that nonlinearity has a bidirectionality. Indeed, the hypothesis of linearity is crucially one of time-symmetry, that is, similarity in fluctuations forwards as backward \cite{theiler1992testing}. It is imaginable that multifractal fluctuations could break time-symmetry in at least two ways in simplest terms, for example, with continuous growth of temporal correlations across time or with continuous decay of temporal correlations. The issue for multifractal tests for nonlinear is that the surrogate specifies the usually nonzero range of variability in temporal correlations. Hence, we recognize nonlinearity as the variability beyond what strictly linear correlations might entail with the shape of the measured series' histogram. Nonetheless, the contrast is bidirectional, like any $t$-test can be bidirectional. Certainly, suppose the original multifractality exceeds a sample of surrogates' multifractality; in that case, nonlinear correlations may be responsible for the excess of multifractality compared to the expectation from strictly linear correlations. Then again, although nonlinearity can amplify heterogeneity, there is no reason the nonlinearity can not constrict and constrain behavior \cite{kelso1995dynamic,lorenz1963deterministic}. 

The present work clarifies and underscores the prior understanding that this greater multifractal spectrum width is a systematic consequence of repeated multiplicative interactions across scales \cite{mandelbrot1974intermittent,halsey1986fractal}. But then we also found that negative $t_\mathrm{MF}$ is equally indicative of random multiplicative cascades, and the major difference between positive and negative $t_\mathrm{MF}$ is that they reflect progressively more and progressively fewer generations of multiplicative interactions. We specifically found that the size of negative $t_\mathrm{MF}$ and the likelihood of significantly negative $t_\mathrm{MF}$ was higher for random multiplicative cascades with fewer generations. Furthermore, negative $t_\mathrm{MF}$ became less prevalent for random multiplicative cascades over progressively more generations. Hence, there is no clear incompatibility between random multiplicative cascades and negative $t_\mathrm{MF}$, and the latter may well be a sign of a less-developed random multiplicative cascade. 

The issue of negative $t_\mathrm{MF}$ does, however, hold further nuance that points us to a novel understanding of how ergodicity breaking might help characterize cascade dynamics correctly. Specifically, the negative $t_\mathrm{MF}$ is not unambiguous. Random multiplicative cascades are likelier to exhibit negative $t_\mathrm{MF}$. Then, it is equally true that successive generations of random \textit{additive} cascades can increase the likelihood of negative $t_\mathrm{MF}$. The inclusion of progressively weakened correlations in noise (e.g., as in our \textit{defractalizing} cascades of both random multiplicative and random additive types) can increase $t_\mathrm{MF}$. Thus, there is a theoretical blindspot for $t_\mathrm{MF}$ in which it is possible that significantly negative $t_\mathrm{MF}$ could reflect random additive cascades or specific noise types. So, we must have a way to distinguish the difference. Fortunately, the sequence-driven ergodicity breaking of these cascade models distinguishes random additive from random multiplicative cascades even when holding length constant. Hence, we might be able to diagnose and estimate the relative age (i.e., in generations) of random multiplicative cascades. Significant positive or significant negative $t_\mathrm{MF}$ with scant sequence-driven ergodicity breaking would reflect older, more developed, or younger, less developed random multiplicative cascades. Meanwhile, significant negative $t$ and strong sequence-driven ergodicity breaking would reflect additive cascades potentially with progressively more uncorrelated noise. 

Further elaborating on the usefulness of considering ergodicity breaking in characterizing cascade processes, the present results indicated a relationship between ergodicity-breaking constituent noise and ergodicity breaking in the resulting cascades. This relationship was clearest in random multiplicative cascades because random additive cascades appeared to break ergodicity to a similar degree across all noise types. The multiplicative interactions across scales appeared to amplify the ergodicity-breaking quality of the constituent noise processes across short to medium scales. Because $fGn$ was more likely than $awGn$ to break ergodicity \cite{deng2009ergodic,mangalam2021point,mangalam2022ergodic}, we saw the strongest evidence of ergodicity breaking over short-to-medium scales in the random multiplicative cascades implicating $fGn$, as well as in the random multiplicative \textit{fractalizing} noise cascades which featured progressively stronger correlations as appearing in $fGn$ over longer generations and the random multiplicative \textit{mixed} noise cascades featuring contributions of $fGn$ in even just $50\%$ of the generations. Hence, although we may expect that random multiplicative cascades have only weak ergodicity breaking, it appears multiplicative interactions across scales allow the ergodicity breaking of the constituent noise to contribute to the ergodicity breaking of cascades.

The advantages conferred by $t_\mathrm{MF}$ and ergodicity breaking discussed above make the present findings amenable to comparison of contexts and studies following different data collection and processing procedures. Further investigation of these choices in different types of real-world data will enhance our ability to reject the null hypothesis of linearity and quantify the strength of the evidence of nonlinear interactions across scales due to multiplicative cascade dynamics characterizing the emergence and creativity of biological and psychological behavior.\\

\noindent{}\textbf{Acknowledgments:} This work was supported by the Center for Research in Human Movement Variability at the University of Nebraska at Omaha, funded by the NIH award P20GM109090.\\

\noindent{}\textbf{Competing interests:} The authors have no conflict of interest.\\

\bibliography{apssamp}

\end{document}